\documentclass[11pt]{article}
\usepackage{latexsym,amssymb} 
\newcommand{\ft}[2]{{\textstyle\frac{#1}{#2}}}

\def\bfone{\relax{\rm 1\kern-.35em 1}}

\newcommand{\llceil}{{|\!\!|\!\!\lceil}}
\newcommand{\rrfloor}{{\rfloor\!\!|\!\!|}}

\begin{document}
\begin{titlepage}

\begin{flushright}\small 
ITP-UU-08/01 \\ SPIN-08/01\\AEI-2007-176\\ENSL-00203073 
\end{flushright}
%%%%%%%%%%%%%%%%%%%%%%%%%%%%%%%%%%%%%%%%%%%%%%%%%%%%%%%%%%%%%%%
%%%%%%%%%%%%%%%%%%%%%%%%%%%%%%%%%%%%%%%%%%%%%%%%%%%%%%%%%%%%%%%
%
\vskip 10mm
\begin{center}
  {\Large {\bf GAUGED SUPERGRAVITIES, TENSOR \\[1.3ex]
      HIERARCHIES, AND {\LARGE M-}THEORY}}
\end{center}
\vskip 8mm

\begin{center}
{{\bf Bernard de Wit}\\
 Institute for Theoretical Physics \,\&\, Spinoza Institute,\\  
Utrecht University, Postbus 80.195, NL-3508 TD Utrecht, 
The Netherlands\\
{\tt b.dewit@phys.uu.nl}}
\vskip 2mm
{{\bf Hermann Nicolai}\\
Max-Planck-Institut f\"ur Gravitationsphysik (Albert-Einstein-Institut),\\
M\"uhlenberg 1, D-14476 Potsdam, Germany\\
{\tt Hermann.Nicolai@aei.mpg.de}}
\vskip 2mm
{{\bf Henning Samtleben}\\
Universit\'e de Lyon, Laboratoire de Physique, \\
Ecole Normale Sup\'erieure de Lyon,\\ 
46 all\'ee d'Italie, F-69364 Lyon CEDEX 07, France \\
{\tt henning.samtleben@ens-lyon.fr}}
\vskip 4mm
\end{center}

\vskip .2in

\begin{center} {\bf Abstract } 
\end{center}
%%%%%%%%%
\begin{quotation}\noindent
  Deformations of maximal supergravity theories induced by gauging
  non-abelian subgroups of the duality group reveal the presence of
  charged M-theory degrees of freedom that are not necessarily
  contained in supergravity. The relation with M-theory degrees of
  freedom is confirmed by the representation assignments under the
  duality group of the gauge charges and the ensuing vector and tensor
  gauge fields. The underlying hierarchy of these gauge fields is
  required for consistency of general gaugings. As an example gauged
  maximal supergravity in three space-time dimensions is presented in
  a version where all possible tensor fields appear. 
\end{quotation}
%%%%%%%%
\end{titlepage}
\eject  
%%%%%%%%%%%%%%%%%%%%%%%%%%%%%%%%%%%%%%%%%%%%%%%%%%%%%%%%%%%%
%%%%%%%%%%%%%%%%%%%%%%%%%%%%%%%%%%%%%%%%%%%%%%%%%%%%%%%%%%%%
\section{Introduction}
\setcounter{equation}{0}
\label{sec:introduction}
%%%%%%%%%%%%%%%%%%%%%%%%%%%%%%%%%%%%%%%%%%%%%%%%%%%%%%%%%%%%

In recent years a wealth of information has become available about
general gaugings of supergravity. In particular, it has become clear
that these theories may play an essential role in probing and exploring 
M-theory beyond the supergravity approximations considered so far.
The key ingredient in these developments is the so-called {\em embedding
tensor}, which defines the embedding of the gauge group generators (up 
to possible central extensions) in the rigid symmetry group, which for
the maximal supergravities is the duality group that arises upon 
dimensional reduction of eleven-dimensional or ten-dimensional 
(IIB) supergravity. With the embedding tensor at hand, all gauged 
supergravities in various space-time dimensions can now be classified.

The first maximal gauged supergravity, $N=8$ supergravity in four
space-time dimensions with compact gauge group $\mathrm{SO}(8)$, was
constructed in \cite{deWit:1981}, soon followed by similar gaugings in
maximal supergravity in $D=5$ \cite{Gunaydin:1984qu} and $D=7$
\cite{Pernici:1984xx} dimensions.  Also $D=4$ gaugings with
non-compact versions of $\mathrm{SO}(8)$ and contractions thereof were
found to exist \cite{Hull}.  Although these results eluded a more
systematic understanding for a long time, there were hints of a deeper
group-theoretical structure underlying these constructions, and
linking the existence of gauged supergravities to certain
higher-dimensional representations of the duality groups
$\mathrm{E}_{n(n)}$: it was known already in 1984 that the so-called
$T$-tensor of $N=8$ supergravity (essentially a `dressed' version of
the embedding tensor) belongs to the $\bf{912}$ representation of
$\mathrm{E}_{7(7)}$ \cite{deWit:1983gs}. The latter group is the
invariance group of (ungauged) maximal supergravity in $D=4$
dimensions \cite{cremmer}.

The more recent developments allowing for a much more systematic
exploration of gauged supergravities go back to the discovery of
maximal gauged supergravities in {\em three} space-time dimensions
\cite{Nicolai:2000sc,Nicolai:2001sv,deWit:2004yr}, and it was in this
context that the notion of embedding tensor was first introduced.  The
case of three space-time dimensions is special because all higher-rank
tensor fields present in higher dimensions have been dualized away in
the dimensional reduction of $D=11$ supergravity \cite{CJS} to three
dimensions, such that all propagating degrees of freedom can be
described by scalar fields. An immediate puzzle then (and the reason
why these theories had not been found earlier) was the question how to
gauge a theory that apparently does not have any vector fields left
from the dimensional reduction. This puzzle was finally resolved in
\cite{Nicolai:2000sc,Nicolai:2001sv} by introducing a set of 248
`redundant' vector fields transforming in the adjoint representation
of $\mathrm{E}_{8(8)}$; rather than through the standard Yang-Mills
kinetic term, these gauge fields appear with a Chern-Simons term in
the Lagrangian, ensuring that the number of physical degrees of
freedom in the theory remains the same as before. 

The systematic investigation of gauged supergravities in dimensions
$D\geq 4$ by means of the embedding tensor was initiated in refs.
\cite{deWit:2002vt,deWit:2003hr}, following the discovery of a new
maximal gauged supergravity in \cite{AndDauFerrLle} based on
Scherk--Schwarz compactification \cite{Scherk:1979zr} of $D=5$
supergravity. This systematic analysis has meanwhile led to a complete
classification of gauged maximal supergravities in $D=5$
\cite{deWit:2004nw}, $D=7$ \cite{Samtleben:2005bp}, and, finally,
$D=4$ \cite{deWit:2007mt} and $D=6$ \cite{Samt} (the situation in an
even number of dimensions is more complicated because the duality
group is only a symmetry of the equations of motion, but not of the
Lagrangian). In particular, it can be shown that the known examples of
gauged supergravities (including more recent constructions such as
\cite{Hull2,deWit:2003hq,Dall'Agata:2005ff,Andrianopoli:2005jv,D'Auria:2005,Schon:2006kz,Hull:2006tp,deVroome:2007zd,Dall'Agata:2007sr,Derendinger:2007xp})
can all be accommodated within the systematic approach based on the
embedding tensor. Most recently, gaugings of maximal supergravity in
$D=2$ were constructed in \cite{SamtlebenWeidner2} --- this case being
more exotic because the relevant duality group $\mathrm{E}_9$ is
infinite dimensional.

The appearance of `redundant' vector fields in $D=3$ gauged
supergravities and the (long known) fact that the consistent gauging
of maximal supergravity in $D=5$ \cite{Gunaydin:1984qu} requires the
simultaneous use of vector fields and 2-form potentials, has led to
the conclusion that a systematic understanding of gauged
supergravities makes the consideration of higher-rank tensor fields
unavoidable \cite{deWit:2005hv}. As pointed out there, and as will be
analyzed in further detail in the present paper, gauged supergravities
can be consistently and systematically formulated by introducing {\em
  a hierarchy of anti-symmetric tensor fields}. The analysis at this
point is independent of the number of space-time dimensions, and the
hierarchy contains in principle an infinite number of anti-symmetric
tensors of any rank. Of course, once the space-time dimension is fixed
to some integer $D$, the maximal rank is also fixed to $D$.
Maintaining the correct number of propagating degrees of freedom in
the presence of these extra fields requires a subtle interplay of
ordinary gauge invariance and higher-rank tensor gauge
transformations. For non-zero gauge coupling the physical degrees of
freedom reside in a finite number of the tensor fields and it is the
embedding tensor that determines how these degrees of freedom are
distributed over the various tensor fields.  Here it is important to
note that, in the presence of gauge interactions, the possibility for
converting rank-$p$ to rank-$(D-p-2)$ tensors fields is severely
restricted. When the gauge coupling constant vanishes the hierarchy
can in general be truncated.

In an important and independent development
\cite{West:2001as,Riccioni:2007au,Bergshoeff:2007qi,Bergshoeff:2007vb},
following earlier papers \cite{West:2001as,Schnakenburg:2001ya}, it
has been shown that the relevant representations of all higher-rank
tensors fields can also be obtained via a level decomposition of the
indefinite Kac--Moody algebra $\mathrm{E}_{11}$ (if one omits the
$D$-forms, these representations can be equivalently derived from the
hyperbolic Kac--Moody algebra $\mathrm{E}_{10}$). In order to arrive
at this decomposition, one first selects a `disabled' node in the
Dynkin diagram, and then decomposes the algebra in representations of
the remaining finite dimensional subgroups of $\mathrm{E}_{11}$, all
of which are direct products $\mathrm{SL}(D) \times \mathrm{E}_{11-D}$
where $D\geq3$ denotes the number of uncompactified space-time
coordinates.  Remarkably, it turns out that the low-lying
representations in that analysis coincide with the representations
found here by a completely different route. However, one should keep
in mind that `higher up' in the level decompositions of
$\mathrm{E}_{10}$ and $\mathrm{E}_{11}$ there opens up a {\em terra
  incognita} of an exponentially growing spectrum of representations
of ever increasing size and complexity, whose ultimate role and
significance remain to be understood.

What is the physical significance of these results? As we will argue
here, the existence of these gauged supergravities constitutes direct
evidence for new M-theoretic degrees of freedom beyond the known
maximal supergravities in space-time dimensions $D\leq 11$ (and
possibly also beyond string theory as presently understood). This
feature is most evident for $D=3$ gauged supergravities with
semi-simple gauge groups: none of these theories can be obtained from
higher-dimensional supergravity by conventional (Kaluza--Klein or
Scherk--Schwarz) compactification.  Our claim is supported by the fact
that several of the `exotic' representations of the duality groups
exhibited here have also been found to occur in toroidally
compactified matrix theory \cite{Elitzur:1997zn,Obers:1998fb}, as well
as in the context of del Pezzo surfaces and compactified M-theory
\cite{Iqbal:2001ye}. The process of gauging a given maximal
supergravity can thus be interpreted as the process of `switching on'
such new degrees of freedom, which are here encoded into the embedding
tensor. A special role is played by the $(D\!-\!1)$- and $D$-forms: we
will set up a Lagrangian formulation of three-dimensional maximal
supergravity containing all higher-rank antisymmetric tensor fields
with an initially space-time dependent embedding tensor $\Theta(x)$,
in such a way that the $(D\!-\!1)$- and $D$-forms, respectively, impose
the constancy of $\Theta$, and the closure of the corresponding gauge
group.  Alternatively, one can eliminate the field $\Theta$ which
appears at most quadratically in the Lagrangian by means of its
equations of motion, thereby arriving at a Lagrangian that contains
the higher-rank tensor fields in a non-polynomial fashion. Gauging
would then be realized as a kind of spontaneous symmetry
breaking,\footnote{%%%%%%%%%%%%%%%%%%%%%%%%%%%%%%%%%%%%%%%%%%%%
  This terminology clearly differs from the usual one, and should thus
  be understood {\em cum grano salis}.} %%%%%%%%%%%%%%%%%%%%%%%
and equivalent to the process of certain $D$-form field strengths
acquiring vacuum expectation values.  In this way, the different
maximal gauged supergravities can be interpreted as different `phases'
of one and the same Lagrangian theory.

Finally, we should stress that we consider the deformations mainly
from the point of view of setting up a consistent gauging. On the
other hand, additional deformations are sometimes possible, generated
by singlet components in the `descendants' of the embedding tensor
(which, presumably, could induce additional non-singlet terms higher
up in the hierarchy).  The embedding tensor by definition specifies
how the gauge group is embedded in the duality group, but it also
encodes many of the interactions of the tensor fields. At the level of
these tensor interactions the embedding tensor may be able to
accomodate additional components which will still fit into the
hierarchy. A well-known example of this phenomenon is the Romans
massive deformation of ten-dimensional IIA supergravity
\cite{Romans:1985tz}, which is induced by a nine-form potential. We
will comment on this in due course.

This paper is organized as follows. In section \ref{sec:hierarchy} we
discuss the hierarchy of tensor gauge fields in a general context.  In
section \ref{sec:m-theory-degrees} we discuss the relation with
M-theory degrees of freedom.  In section \ref{sec:three-dimensions} we
determine the duality representations of the tensor fields in three
space-time dimensions. The corresponding supersymmetry algebra is
discussed in section~\ref{sec:susy-algebra} and the general Lagrangian
for gauged three-dimensional maximal supergravity in
section~\ref{sec:Lagrangian-forms-3d}.  Results of the present
investigation have already been announced and discussed by us in
several talks~\footnote{See, for instance:
  {http://ggi-www.fi.infn.it/activities/workshops/stringM/talks/dewit.pdf};
  \\
{http://maths.dur.ac.uk/events/Meetings/LMS/2007/TSAS/Talks/dewit.pdf}
}.

%%%%%%%%%%%%%%%%%%%%%%%%%%%%%%%%%%%%%%%%%%%%%%%%%%%%%%%%%%%
%%%%%%%%%%%%%%%%%%%%%%%%%%%%%%%%%%%%%%%%%%%%%%%%%%%%%%%%%%%
\section{A hierarchy of vector and tensor gauge fields}
\setcounter{equation}{0}
\label{sec:hierarchy}
%%%%%%%%%%%%%%%%%%%%%%%%%%%%%%%%%%%%%%%%%%%%%%%%%%%%%%%%%%%%
Maximal supergravities in various space-time dimensions can be
constructed by dimensional reduction on a torus of supergravity in
eleven and/or ten space-time dimensions. In general these theories
contain abelian vector fields and antisymmetric tensor fields of
various ranks. Their field content is not unique as $p$-rank tensor
gauge fields can be dualized to tensor fields of rank $D-p-2$, where
$D$ denotes the dimension of space-time of the reduced theory.
However, there always exists an optimal choice of the field
configuration that most clearly exhibits the invariance under a
duality group $\mathrm{G}$.  This group is listed for space-time
dimensions $D= 3,\ldots,7$ in the second column of table
\ref{tab:T-tensor-repr}. The symmetry under the
$\mathrm{G}$-transformations is realized non-linearly in view of the
fact that the scalar fields parametrize a $\mathrm{G}/\mathrm{H}$
coset space, where $\mathrm{H}$ is the R-symmetry group of the
corrresponding supersymmetry algebra.  This group equals the maximal
compact subgroup of $\mathrm{G}$ and it is also listed in
table~\ref{tab:T-tensor-repr}.  In general the vector and
antisymmetric gauge fields transform in specific representations of
$\mathrm{G}$.\footnote{%%%%%%%%%%%%%%%%%%%%%%%%%%%%%%%%%%%%%%%%
  In even space-time dimensions this assignment may fail and complete
  $\mathrm{G}$ representations may require the presence of magnetic
  duals. For four space-time dimensions, this has been demonstrated in
  \cite{deWit:2005ub}.}
%%%%%%%%%%%%%%%%%%%%%%%%%%%
The vector fields, which we denote by $A_\mu{}^\mathcal{M}$, transform
in the fundamental or in a spinor representation of~$\mathrm{G}$.
These representations are (implicitly) listed in table
\ref{tab:T-tensor-repr}, as we will explain below.  The generators in
these representations are denoted by
$(t_\alpha)_{\mathcal{M}}{}^{\mathcal{N}}$, so that $\delta
A_\mu{}^\mathcal{M}=-\Lambda^\alpha(t_\alpha)_\mathcal{N}{}^\mathcal{M}
\,A_\mu{}^\mathcal{N}$. Structure constants $f_{\alpha\beta}{}^\gamma$
of the duality group are defined according to $[t_\alpha,t_\beta]=
f_{\alpha\beta}{}^\gamma \,t_\gamma$.

Deformations of these maximal supergravities can be constructed by
introducing a non-abelian gauge group, which must be a subgroup of the
duality group. The dimension of this gauge group is obviously
restricted by the number of vector fields in the theory. The
discussion in this section will remain rather general and will neither
depend on the actual duality group nor on the space-time dimension (we
recall, however, that there may be subtleties in even space-time
dimensions related to selfduality of vector or tensor gauge
fields). We refer to \cite{deWit:2004nw,Samtleben:2005bp,deWit:2007mt} 
where a number of results were described for maximal supergravity in
various dimensions. 

The gauge group generators $X_\mathcal{M}$, which will couple to the gauge
fields $A_\mu{}^\mathcal{M}$ in the usual fashion, are obviously
decomposed in terms of the independent $\mathrm{G}$ generators
$t_\alpha$, {\it i.e.},
\begin{equation}
  \label{eq:X-theta-t}
  X_\mathcal{M} = \Theta_\mathcal{M}{}^\alpha\,t_\alpha \;.
\end{equation}
The gauging is thus encoded in a real {\it embedding tensor}
$\Theta_\mathcal{M}{}^{\alpha}$ belonging to the product of the
representation conjugate to the representation in which the gauge
fields transform and the adjoint representation of $\mathrm{G}$. This
product representation is reducible and decomposes into a number of
irreducible representations as is indicated for the cases of interest
in the last column of table~\ref{tab:T-tensor-repr}.  However, as is
also shown in the table, supersymmetry requires most of these
irreducible representations to be absent: only the underlined
representations in the table are compatible with local supersymmetry.
Actually, for non-supersymmetric theories one may have to impose
similar constraints (see, {\it e.g.}  \cite{deWit:2005ub}). This
constraint on the embedding tensor is known as the {\it representation
  constraint}. Here we treat the embedding tensor as a spurionic
object, which we allow to transform under the duality group so that
the Lagrangian and transformation rules remain formally invariant
under $\mathrm{G}$. At the end we will freeze the embedding tensor to
a constant, so that the duality invariance will be broken.  Later in
this paper we see that this last step can also be described in terms
of a new action in which the freezing of $\Theta_\mathcal{M}{}^\alpha$
will be the result of a more dynamical process.

%%%%%%%%%%%%%%%%%%%%%%%%%%%%%%%%%%%%%%%%%%%%%%%%%%%%%%%%%%%%%%%%%%
%%%%%%%%%%%%%%%%%%%%%%%%%%%%%%%%%%%%%%%%%%%%%%%%%%%%%%%%%%%%%%%%%%
\begin{table}[t]
\centering
\begin{tabular}{l l  l l  }\hline
~&~&~&~\\[-4mm]
$D$ &${\rm G}$& ${\rm H}$ & $\Theta$  \\   \hline
~&~&~&~\\[-4mm]
7   & ${\rm SL}(5)$ & ${\rm USp}(4)$  & ${\bf 10}\times {\bf
  24}= {\bf 10}+\underline{\bf 15}+  \underline{\overline{\bf 40}}+
{\bf 175}$ 
\\[1mm] 
6  & ${\rm SO}(5,5)$ & ${\rm USp}(4) \times {\rm USp}(4)$ & 
  ${\bf 16}\times{\bf 45} =
  {\bf 16}+ \underline{\bf 144} + {\bf 560}$ \\[.8mm]
5   & ${\rm E}_{6(6)}$ & ${\rm USp}(8)$ & ${\bf 27}\times{\bf
  78} =  
  {\bf 27} + \underline{\bf 351} + \overline{\bf 1728}$  \\[.5mm]
4   & ${\rm E}_{7(7)}$ & ${\rm SU}(8)$  & ${\bf 56}\times{\bf 133} = 
  {\bf 56} + \underline{\bf 912} + {\bf 6480}$   \\[.5mm]
3   & ${\rm E}_{8(8)}$ & ${\rm SO}(16)$ & ${\bf 248}\times{\bf 248} = 
  \underline{\bf 1} + {\bf 248} + \underline{\bf 3875} +{\bf 27000} 
  +  {\bf 30380}$ 
\\ \hline
\end{tabular}
\caption{\small
Decomposition of the embedding tensor $\Theta$ for maximal
supergravities in various space-time dimensions in terms of irreducible
${\rm G}$ representations 
\cite{deWit:2002vt,deWit:2005hv}. Only the underlined representations
are allowed by supersymmetry. The R-symmetry group ${\rm H}$
is the maximal compact subgroup of ${\rm G}$.  
}\label{tab:T-tensor-repr}
\end{table}
%%%%%%%%%%%%%%%%%%%%%%%%%%%%%%%%%%%%%%%%%%%%%%%%%%%%%%%%%%%%%%%%%%
%%%%%%%%%%%%%%%%%%%%%%%%%%%%%%%%%%%%%%%%%%%%%%%%%%%%%%%%%%%%%%%%%%

The embedding tensor must satisfy a second constraint, the so-called
{\em closure constraint}, which is quadratic in
$\Theta_\mathcal{M}{}^\alpha$ and more generic. This constraint
ensures that the gauge transformations form a group so that the
generators (\ref{eq:X-theta-t}) will close under commutation. Any
embedding tensor that satisfies the closure constraint, together with
the representation constraint mentioned earlier, defines a consistent
gauged supergravity theory that is both supersymmetric and gauge
invariant. To spell out the closure constraint in more detail let us
write out (\ref{eq:X-theta-t}) once more, but now with representation
indices in the G-representation pertaining to the gauge fields written
out explicitly, {\it viz.}
\begin{equation}
  \label{eq:X}
  X_{\mathcal M \mathcal N}{}^{\mathcal P} \equiv
  \Theta_\mathcal{M}{}^\alpha\,(t_\alpha)_{\mathcal N}{}^{\mathcal P} 
  = X_{[\mathcal M \mathcal N]}{}^{\mathcal P} 
  +  Z^\mathcal{P}{}_{\mathcal{MN}}
\end{equation}
where we will use the notation
\begin{equation}
  \label{eq:def-Z}
  Z^\mathcal{P}{}_{\mathcal{MN}} \equiv
  X_{(\mathcal{MN})}{}^\mathcal{P} \,,
\end{equation}
for the symmetric part throughout this paper. The closure constraint
is a consequence of the invariance of the embedding tensor under the
gauge group it generates, that is
\begin{equation}
  \label{eq:gauge-inv-embedding}
  \delta_{\mathcal P} \Theta_{\mathcal M}{}^\alpha 
  = \Theta_{\mathcal P}{}^\beta t_{\beta \mathcal{M}}{}^{\mathcal N}
    \Theta_{\mathcal N}{}^\alpha +
    \Theta_{\mathcal P}{}^\beta f_{\beta\gamma}{}^\alpha 
    \Theta_{\mathcal M}{}^\gamma = 0 \,.
\end{equation}
Contracting this result with $t_\alpha$ we obtain
\begin{equation}
  \label{eq:XX-commutator}
{[X_\mathcal{M},X_\mathcal{N}]} =
-X_\mathcal{MN}{}^\mathcal{P}\,X_\mathcal{P} = 
-X_{[\mathcal{MN}]}{}^\mathcal{P}\,X_\mathcal{P}    \,, 
\end{equation}
Hence, the gauge invariance of the embedding tensor is equivalent to
the closure of the gauge algebra.
%gauge invariance of the tensor $X_{\mathcal{MN}}{}^\mathcal{P}$. 
It is noteworthy here that the generator
$X_\mathcal{MN}{}^\mathcal{P}$ and the structure constants of the
gauge group are thus related, but do not have to be identical. In
particular $X_\mathcal{MN}{}^\mathcal{P}$ is in general not
antisymmetric in $[\mathcal{MN}]$, as is evident from (\ref{eq:X}).
The embedding tensor acts as a projector, and only in the projected
subspace the matrix $X_\mathcal{MN}{}^\mathcal{P}$ is antisymmetric in
$[\mathcal{MN}]$ and the Jacobi identity will be satisfied. Therefore
(\ref{eq:XX-commutator}) implies in particular that
$X_{(\mathcal{MN})}{}^\mathcal{P}$ must vanish when contracted with
the embedding tensor. In terms of the notation introduced above, this
condition reads
\begin{equation}
  \label{eq:closure}
  \Theta_\mathrm{\mathcal{P}}{}^\alpha \, Z^\mathcal{P}{}_{\mathcal{MN}}
  =0\,. 
\end{equation}
The gauge invariant tensor $Z^\mathcal{P}{}_{\mathcal{MN}}$ transforms
in the same representation as $\Theta_\mathcal{M}{}^\alpha$, except
when the embedding tensor transforms reducibly so that
$Z^\mathcal{P}{}_\mathcal{MN}$ may depend on a smaller representation.
As may be expected the tensor $Z^\mathcal{P}{}_{\mathcal{MN}}$
characterizes the lack of closure of the generators $X_\mathcal{M}$.
This can be seen, for instance, by calculating the direct analogue of
the Jacobi identity, 
\begin{equation}
\label{Jacobi-X}
   X_{[\mathcal{NP}}{}^\mathcal{R}
   \,X_{\mathcal{Q}]\mathcal{R}}{}^\mathcal{M} = 
   \ft23 Z^\mathcal{M}{}_{\mathcal{R}[\mathcal{N}}\,
   X_{\mathcal{PQ}]}{}^\mathcal{R} \,.
\end{equation}

We emphasize that seemingly strange features, such as the appearance
of a symmetric contribution in $X_{\mathcal{MN}}{}^\mathcal{P}$, or
the apparent violation of the Jacobi identity in (\ref{Jacobi-X}), are 
entirely due to the redundancy in the description: although the actual 
gauge group is usually smaller than G, we nevertheless continue to label 
all matrices by G indices $\mathcal M$, such that the number of matrices 
$X_{\mathcal M}$ in general will exceed the dimension of the gauge group.
The main advantage of this parametrization (and nomenclature) is its 
universality, which allows us to treat all gaugings (and gauge groups) 
on the same footing.

Now we return to the field theoretic description. The gauging requires
the replacement of ordinary space-time derivatives by covariant ones
for all fields except the gauge fields,
\begin{equation}
  \label{eq:vector-gauge-tr}
  \partial_\mu\to D_\mu = \partial_\mu -g\, A_\mu{}^\mathcal{M}
  \,X_\mathcal{M} \,, 
\end{equation}
where the generator $X_\mathcal{M}$ must be taken in the appropriate
representation. To write down invariant kinetic terms for the gauge
fields one needs a suitable covariant field strength tensor. This is
an issue because the Jacobi identity is not satisfied. The standard
field strength, which follows from the Ricci identity, $[D_\mu,D_\nu]=
- g \mathcal{F}_{\mu\nu}{}^\mathcal{M}\,X_\mathcal{M}$, reads, 
\begin{equation}
  \label{eq:field-strength}
  \mathcal{F}_{\mu\nu}{}^\mathcal{M} =\partial_\mu A_\nu{}^\mathcal{M}
  -\partial_\nu
  A_\mu{}^\mathcal{M} + g\, X_{[\mathcal{NP}]}{}^\mathcal{M}
  \,A_\mu{}^\mathcal{N} A_\nu{}^\mathcal{P} \,, 
\end{equation}
and is not fully covariant. The lack of covariance can be readily
checked by observing that $\mathcal{F}_{\mu\nu}{}^\mathcal{M}$ does
{\em not} satisfy the Palatini identity~\footnote{That is, the standard  
  relation  $\delta\mathcal{F}_{\mu\nu}{}^\mathcal{M} = 
  2\, D_{[\mu}\delta A_{\nu]}{}^\mathcal{M}$.}; rather, we have 
\begin{equation}
  \label{eq:Palatini}
  \delta\mathcal{F}_{\mu\nu}{}^\mathcal{M} = 2\, D_{[\mu}\delta
  A_{\nu]}{}^\mathcal{M} - 
  2 g\, X^\mathcal{M}{}_{(\mathcal{PQ})} \,A_{[\mu}{}^\mathcal{P}
  \,\delta A_{\nu]}{}^\mathcal{Q} \,,
\end{equation}
under arbitrary variations $\delta A_\mu{}^\mathcal{M}$. Assuming the
standard gauge transformation,
\begin{equation}
  \label{eq:A-var}
  \delta A_\mu{}^\mathcal{M} =  D_\mu\Lambda^\mathcal{M} =
  \partial_\mu \Lambda^\mathcal{M} + g A_\mu{}^\mathcal{N}
  X_\mathcal{NP}{}^\mathcal{M} \Lambda^\mathcal{P} \,,   
\end{equation}
it follows that $\mathcal{F}_{\mu\nu}{}^M$ transforms under gauge
transformations as
\begin{equation}
  \label{eq:delta-cal-F}
  \delta\mathcal{F}_{\mu\nu}{}^\mathcal{M}= g\, \Lambda^\mathcal{P}
  X_\mathcal{NP}{}^\mathcal{M}
  \,\mathcal{F}_{\mu\nu}{}^\mathcal{N} - 2 g\, Z^\mathcal{M}{}_\mathcal{PQ}
  \,A_{[\mu}{}^\mathcal{P}\,\delta A_{\nu]}{}^\mathcal{Q}  \,, 
\end{equation} 
which is {\em not} covariant --- not only because of the presence of
the second term on the right-hand side, but also because the lack of
antisymmetry of the structure constants $X_\mathcal{NP}{}^\mathcal{M}$
prevents us from getting the correct result (cf.
(\ref{eq:delta-cov-H-2}) below) by simply inverting the order of
indices $\mathcal{NP}$ in the first term on the right-hand side

In order to remedy this lack of covariance we now follow
the strategy of \cite{deWit:2004nw,deWit:2005hv}. Since we know that 
closure is ensured on the subspace projected by the embedding tensor, 
we introduce additional gauge transformations in the orthogonal
complement so that all difficulties associated with the lack of
closure can be compensated for by performing these new
transformations.  For the gauge fields, this leads to the following
transformation rule,
\begin{equation}
  \label{eq:A-var-2}
\delta A_\mu{}^\mathcal{M} =  D_\mu\Lambda^\mathcal{M} -
g\,Z^\mathcal{M}{}_{\mathcal{NP}}\,\Xi_\mu{}^{\mathcal{NP}}\,,   
\end{equation}
where the transformations proportional to $\Xi_\mu{}^\mathcal{NP}$
enable one to gauge away those vector fields that are in the sector of
the gauge generators $X_\mathcal{MN}{}^\mathcal{P}$ where the Jacobi
identity is not satisfied (this sector is perpendicular to the embedding 
tensor by (\ref{eq:closure})). Note that the parameter 
$\Xi_\mu{}^{\mathcal{NP}}$ in (\ref{eq:A-var-2}) appears contracted 
with the constant tensor
$Z^\mathcal{M}{}_{\mathcal{NP}}$ defined in (\ref{eq:def-Z}) as a
linear function of the embedding tensor.  It is important, that this
tensor generically does not map onto the full symmetric tensor product
$(\mathcal{NP})$ in its lower indices but rather only on a restricted
subrepresentation. In other words, there is a non-trivial $\rm
G$-invariant projector~$\mathbb{P}$ such that 
\begin{equation}
  \label{eq:Z-proj}
Z^\mathcal{M}{}_{\mathcal{NP}} = 
Z^\mathcal{M}{}_{\mathcal{RS}}\,{\mathbb
  P}^{\mathcal{RS}}{}_{\mathcal{NP}} \;,  
\end{equation}
for any choice of the embedding tensor. The precise representation
content of $\mathbb{P}$ can be determined for any given theory by
carefully inspecting~(\ref{eq:def-Z}) and we give examples of this in
the later sections (see also~\cite{deWit:2005hv}).  In order not to
overburden the formulas with explicit projectors, we denote the
projection corresponding to~(\ref{eq:Z-proj}) by the special brackets
$\llceil{\mathcal{NP}}\rrfloor$, {\it i.e.}\ we use the notation
\begin{equation}
  \label{eq:bracket-exd5}
 A^{\llceil\cal M}A^{{\cal N}\rrfloor} \equiv {\mathbb
  P}^{\mathcal{MN}}{}_{\mathcal{RS}}\,A^{\cal R}A^{{\cal S}} \;,\qquad
\mbox{etc.}   
\end{equation}
Similar notation will be used for other index combination that we will
encounter shortly. 

The combined gauge transformations (\ref{eq:A-var-2}) generate a group
on the vector fields, as follows from the commutation relations,
\begin{eqnarray}
  \label{eq:gauge-commutators}
  {}[\delta(\Lambda_1),\delta(\Lambda_2)] &=& \delta(\Lambda_3) +
  \delta(\Xi_3) \,, 
\end{eqnarray}
where 
\begin{eqnarray}
  \label{eq:gauge-parameters}
  \Lambda_3{}^\mathcal{M} &=& g\,X_{[\mathcal{NP}]}{}^\mathcal{M} \,
  \Lambda_2^\mathcal{N}\Lambda_1^\mathcal{P}\,, \nonumber\\
  \Xi_{3 \mu}{}^{\mathcal{MN}} &=&  \Lambda_1^{\llceil\mathcal{M}}
  D_\mu\Lambda_2^{\mathcal{N}\rrfloor} -
  \Lambda_2^{\llceil\mathcal{M}} D_\mu \Lambda_1^{\mathcal{N}\rrfloor} 
  \,.
\end{eqnarray}
Here it is crucial that $\delta(\Lambda)$ and $\delta(\Xi)$ commute on
the vector fields. However, these commutators are subject to change
when more fields will be introduced. We return to this issue in
due course.

Under the combined gauge transformations $\mathcal{F}_{\mu\nu}{}^M$
changes as follows,
\begin{equation}
  \label{eq:delta-cal-F-combined}
  \delta\mathcal{F}_{\mu\nu}{}^\mathcal{M}= g\, \Lambda^\mathcal{P}
  X_\mathcal{NP}{}^\mathcal{M} 
  \,\mathcal{F}_{\mu\nu}{}^\mathcal{N} - 2 g\,
  Z^\mathcal{M}{}_\mathcal{PQ} \left(D_{[\mu} \Xi_{\nu]}{}^\mathcal{PQ} +
  \,A_{[\mu}{}^\mathcal{P}\,\delta A_{\nu]}{}^\mathcal{Q}\right) \,, 
\end{equation} 
which is still not covariant. The standard strategy
\cite{deWit:2004nw,deWit:2005hv} is therefore to define modified field
strengths,
\begin{equation}
  \label{eq:modified-fs}
{\cal H}_{\mu\nu}{}^{\mathcal{M}} = 
{\cal F}_{\mu\nu}{}^\mathcal{M}  + g\, Z^\mathcal{M}{}_\mathcal{NP} 
\,B_{\mu\nu}{}^\mathcal{NP}\;, 
\end{equation}
where we introduce tensor fields $B_{\mu\nu}{}^\mathcal{NP}$,
transforming under ${\rm G}$ in the restricted
representation~(\ref{eq:Z-proj}) {\it
  i.e.}~$B_{\mu\nu}{}^\mathcal{NP}=
B_{\mu\nu}{}^\mathcal{\llceil{\mathcal{NP}}\rrfloor}$.  Actually the
restricted index pair $\llceil\mathcal{MN}\rrfloor$ will play the role
of a new index belonging to a specific representation, and
$Z^\mathcal{M}{}_\mathcal{NP}$ is an intertwining tensor between the
representations of the vectors and the two-forms.  The gauge
transformation rules of $B_{\mu\nu}{}^\mathcal{MN}$ will be chosen
such that the field strengths $\mathcal{H}_{\mu\nu}{}^\mathcal{M}$
will transform covariantly under gauge transformations, {\it i.e.},
\begin{equation}
  \label{eq:delta-cov-H-2}
  \delta\mathcal{H}_{\mu\nu}{}^{\mathcal{M}}= 
  - g\, \Lambda^\mathcal{P} X_\mathcal{PN}{}^\mathcal{M} 
  \,\mathcal{H}_{\mu\nu}{}^{\mathcal{N}} \,.
\end{equation} 

To do this in a systematic manner we first define generic covariant
variations of the tensor fields,
\begin{equation}
  \label{eq:cov-delta-B}
  \Delta B_{\mu\nu}{}^{\mathcal{MN}} \equiv
  \delta B_{\mu\nu}{}^{\mathcal{MN}} - 2\, A_{[\mu}{}^{\llceil\mathcal{M}}
  \delta A_{\nu]}{}^{\mathcal{N}\rrfloor}  \,,
\end{equation}
so that generic variations of $\mathcal{H}_{\mu\nu}{}^{\mathcal{M}}$
take the form 
\begin{equation}
  \label{eq:var-H-2}
  \delta\mathcal{H}_{\mu\nu}{}^{\mathcal{M}}= 
  2\, D_{[\mu} \delta A_{\nu]}{}^\mathcal{M} + g\,
  Z^\mathcal{M}{}_\mathcal{NP}\, \Delta B_{\mu\nu}{}^\mathcal{NP} \,. 
\end{equation} 
For a combined gauge transformation we choose for $\Delta
B_{\mu\nu}{}^{\mathcal{MN}}$, 
\begin{equation}
  \label{eq:Delta-gauge-B1}
  \Delta B_{\mu\nu}{}^{\mathcal{MN}}\Big\vert_\mathrm{gauge} = 
    2\,D_{[\mu}\Xi_{\nu]}{}^{\mathcal{MN}} -2\, 
  \Lambda^{\llceil\mathcal{M}}\mathcal{H}_{\mu\nu}{}^{\mathcal{N}\rrfloor}
  + \cdots  \,, 
\end{equation}
where the unspecified contributions vanish when $\Delta
B_{\mu\nu}{}^{\mathcal{MN}}$ is contracted with
$Z^\mathcal{P}{}_\mathcal{MN}$, so that they remain as yet
undetermined. Substituting this expression and (\ref{eq:A-var-2}) into
(\ref{eq:var-H-2}) leads indeed to the required result
(\ref{eq:delta-cov-H-2}).\footnote{%%%%%%%%%%%%%%%%%%%%%%%%%%%%%%%%%%%
  Here we note that the present formulae cannot be compared directly
  to the ones in \cite{deWit:2005hv}, as those are derived in a
  different basis, but they can be compared to later work along the
  same lines, starting with \cite{Samtleben:2005bp}.
} %%%%%%%%%%%%%%%%%%%%%%%%%%%%%%%%%%%%%%%%%%%%%%%%%

Here it is worth pointing out that the expected gauge transformation
on $B_{\mu\nu}{}^\mathcal{MN}$ equal to 
\begin{equation}
  \label{eq:standard-delta-B}
  \delta B_{\mu\nu}{}^\mathcal{MN}= -g\Lambda^\mathcal{P}
  X_{\mathcal{P}\llceil\mathcal{RS}\rrfloor}{}^{\llceil\mathcal{MN}\rrfloor}
  B_{\mu\nu}{}^\mathcal{RS} \,,
\end{equation}
where the generator $X_{\mathcal{P}\llceil\mathcal{RS}\rrfloor}
{}^{\llceil\mathcal{MN}\rrfloor} =
(X_\mathcal{P})_{\llceil\mathcal{RS}\rrfloor} 
{}^{\llceil\mathcal{MN}\rrfloor}$ acts in the restricted
representation to which $\delta B_{\mu\nu}{}^\mathcal{MN}$ belongs,
is already contained in the second term in (\ref{eq:Delta-gauge-B1}),
up to an additional gauge transformation associated with a three-rank
tensor field, that we will introduce shortly.

The above strategy forms the starting point for the construction of a
hierarchy of antisymmetric tensor gauge fields \cite{deWit:2005hv}. To
see how one proceeds, let us turn to the construction of the covariant
field strength for the tensor fields
$B_{\mu\nu}{}^{\mathcal{MN}}$,\footnote{%%%%%%%%%%%%%%%%%%%%%%%%%  
  We use the same letters $\mathcal{F}$ for the field strengths of
  vectors and higher $p$-forms. From the number of space-time indices
  it is always clear to which forms the $\mathcal{F}$ belong. } 
%%%%%%%%%%%%%%%%%%%%%%%%%%%%%%%%%%%%%%%%%%%%%%%%%%%%%%%%%%%%%%%%%%
\begin{equation}
  \label{eq:F-3}
  \mathcal{F}_{\mu\nu\rho}{}^\mathcal{MN} = 
  3\, D_{[\mu} B_{\nu\rho]}{}^\mathcal{MN}  +
  6\,A_{[\mu}{}^{\llceil\mathcal{M}}\left(\partial_{\nu}
  A_{\rho]}{}^{\mathcal{N}\rrfloor} + \ft13 g
  X_{[\mathcal{PQ}]}{}^{\mathcal{N}\rrfloor}
  A_{\nu}{}^\mathcal{P}A_{\rho]}{}^\mathcal{Q}\right)  \,,
\end{equation}
where the first two coefficients follow from~(\ref{eq:Delta-gauge-B1})
and the terms cubic in the vector gauge fields are such that generic
variations of $\mathcal{F}_{\mu\nu\rho}{}^{\mathcal{MN}}$ read as
follows,
\begin{eqnarray}
  \label{eq:var-F-3}
  \delta\mathcal{F}_{\mu\nu\rho}{}^{\mathcal{MN}} &=&
  3\,D_{[\mu}\,\Delta B_{\nu\rho]}{}^\mathcal{MN} + 6\,
  \mathcal{H}_{[\mu\nu}{}^{\llceil\mathcal{M}}
  \,\delta A_{\rho]}{}^{\mathcal{N}\rrfloor} \nonumber\\[.5ex]
 &&{}
 - g\,
  Y^{\mathcal{MN}}{}_{\mathcal{P}\llceil\mathcal{RS}\rrfloor}
  \,(3\, B_{[\mu\nu}{}^\mathcal{RS} \,\delta A_{\rho]}{}^\mathcal{P}
  + 2\, A_{[\mu}{}^\mathcal{P} A_\nu{}^\mathcal{R}\delta
  A_{\rho]}{}^\mathcal{S} )\,, 
\end{eqnarray}
where 
\begin{equation}
  \label{eq:def-Y}
  Y^\mathcal{MN}{}_{\mathcal{P}\llceil\mathcal{RS}\rrfloor} =
 2\,\delta_\mathcal{P}{}^{\llceil\mathcal{M}}
  \,Z^{\mathcal{N}\rrfloor}{}_\mathcal{RS} 
  - X_{\mathcal{P}\llceil\mathcal{RS}\rrfloor}{}^{\llceil\mathcal{MN}\rrfloor}
\;.
\end{equation}
%and
%$X_{\mathcal{P}\llceil\mathcal{RS}\rrfloor}{}^{\llceil\mathcal{MN}\rrfloor}$
%denotes the gauge generator in the restricted subrepresentation
%$\llceil\mathcal{MN}\rrfloor$. 
Note that this definition can be rewritten as
\begin{equation}
  \label{eq:def-Y-2}
  Y^\mathcal{MN}{}_{\mathcal{P}\llceil\mathcal{RS}\rrfloor} =
 2 \left(\delta_\mathcal{P}{}^{\llceil\mathcal{M}}
  \,Z^{\mathcal{N}\rrfloor}{}_\mathcal{RS} 
  -  X_{\mathcal{P}\llceil\mathcal{R}}{}^{\llceil\mathcal{M}}
 \delta_{\mathcal{S}\rrfloor}{}^{\mathcal{N}\rrfloor}\right) \;.
\end{equation}
Just as before we introduce an extra gauge invariance to eventually
deal with the non-covariant variations in the last term of
(\ref{eq:var-F-3}), which will then provide the missing variations in
(\ref{eq:Delta-gauge-B1}),
\begin{equation}
  \label{eq:Delta-gauge-B2}
  \Delta B_{\mu\nu}{}^{\mathcal{MN}}\Big\vert_\mathrm{gauge} = 
    2\,D_{[\mu}\Xi_{\nu]}{}^{\mathcal{MN}} -2\, 
  \Lambda^{\llceil\mathcal{M}}\mathcal{H}_{\mu\nu}{}^{\mathcal{N}\rrfloor}
   - g\, Y^\mathcal{MN}{}_{\mathcal{P}\llceil\mathcal{RS}\rrfloor}
   \Phi_{\mu\nu}{}^{\mathcal{P}\llceil\mathcal{RS}\rrfloor} \,, 
\end{equation}
where $\Phi_{\mu\nu}{}^{\mathcal{P}\llceil\mathcal{RS}\rrfloor}$ is the
new gauge parameter. Secondly we introduce a corresponding three-form
gauge field $C_{\mu\nu\rho}{}^{\mathcal{P}\llceil\mathcal{RS}\rrfloor}$,
and define the field strength $\mathcal{H}_{\mu\nu\rho}{}^{\mathcal{MN}}$, 
\begin{equation}
  \label{eq:H-3-cov}
  \mathcal{H}_{\mu\nu\rho}{}^{\mathcal{MN}}=
  \mathcal{F}_{\mu\nu\rho}{}^{\mathcal{MN}}  + g\,
  Y^\mathcal{MN}{}_{\mathcal{P}\llceil\mathcal{RS}\rrfloor}
  \,C_{\mu\nu\rho}{}^{\mathcal{P}\llceil\mathcal{RS}\rrfloor} \,. 
\end{equation}
such that it transforms covariantly, {\it i.e.}
\begin{equation}
 \delta \mathcal{H}_{\mu\nu\rho}{}^{\mathcal{MN}}= - g \Lambda^{\mathcal P} 
  X_{\mathcal{P}\llceil\mathcal{RS}\rrfloor}{}^{\llceil\mathcal{MN}\rrfloor}
  \, \mathcal{H}_{\mu\nu\rho}{}^{\mathcal{RS}}\,, 
\end{equation}
in complete analogy with (\ref{eq:delta-cov-H-2}). As before, the
tensor $ Y^\mathcal{MN}{}_{\mathcal{P}\llceil\mathcal{RS}\rrfloor}$
does not map onto the full tensor product
${\mathcal{P}\llceil\mathcal{RS}\rrfloor}$ in its lower indices but
only on a restricted subrepresentation inside, {\it i.e.},
\begin{eqnarray}
Y^\mathcal{MN}{}_{\mathcal{P}\llceil\mathcal{RS}\rrfloor}&=&
Y^\mathcal{MN}{}_{\mathcal{Q}\llceil\mathcal{KL}\rrfloor}\,
{\mathbb{P}}^{\mathcal{Q}\llceil\mathcal{KL}\rrfloor}
{}_{\mathcal{P}\llceil\mathcal{RS}\rrfloor}\,  
\;,
\label{eq:Y-proj}
\end{eqnarray}
for a non-trivial projector $\mathbb{P}$ independent of the embedding
tensor. In principle, this projector can be worked out from (\ref{eq:def-Y}), 
but deriving more explicit expressions requires a case-by-case 
consideration for each duality group G. As in (\ref{eq:bracket-exd5}) we will 
denote the corresponding projection by special brackets
$\llceil{\mathcal{P}\llceil\mathcal{RS}\rrfloor}\rrfloor$.  The tensor
$Y^\mathcal{MN}{}_{\mathcal{P}\llceil\mathcal{RS}\rrfloor}$ thus
represents an intertwining tensor between the two- and the
three-forms.  It satisfies the properties
\begin{eqnarray}
  \label{eq:Z-Y-orthogonal}
  Z^\mathcal{Q}{}_\mathcal{MN}\, 
  Y^\mathcal{MN}{}_{\mathcal{P}\llceil\mathcal{RS}\rrfloor}&=&0\,,
  %\nonumber
  \\
  \label{eq:ZZ-YZ}
  Z^\mathcal{K}{}_\mathcal{PQ}\, 
  Y^\mathcal{MN}{}_{\mathcal{K}\llceil\mathcal{RS}\rrfloor} &=& 2\,
  Z^{\llceil\mathcal{M}}{}_\mathcal{PQ} \,
  Z^{\mathcal{N}\rrfloor}{}_\mathcal{RS} \,. 
\end{eqnarray}
which are both consequences of the quadratic
constraint~(\ref{eq:XX-commutator}). The first identity represents the
analogue of~(\ref{eq:closure}). Another identity follows
directly from (\ref{eq:def-Y-2}), 
\begin{equation}
  \label{eq:Y-symmetrized}
  Y^\mathcal{MN}{}_{\mathcal{P}\llceil\mathcal{RS}\rrfloor}
  \Big\vert_{(\mathcal{PRS})}=0\,,
\end{equation}

Generic variations of the covariant field strength~(\ref{eq:closure})
can be written as
\begin{equation}
  \label{eq:var-H-3}
  \delta\mathcal{H}_{\mu\nu\rho}{}^{\mathcal{MN}} =
  3\,D_{[\mu}\,\Delta B_{\nu\rho]}{}^\mathcal{MN} + 6\,
  \mathcal{H}_{[\mu\nu}{}^{\llceil\mathcal{M}}
  \,\delta A_{\rho]}{}^{\mathcal{N}\rrfloor}  + g\,
  Y^{\mathcal{MN}}{}_{\mathcal{P}\llceil\mathcal{RS}\rrfloor}  
  \Delta C_{\mu\nu\rho}{}^{\mathcal{P}\llceil\mathcal{RS}\rrfloor} \,, 
\end{equation}
where
\begin{equation}
  \label{eq:def-cov-C}
  \Delta C_{\mu\nu\rho}{}^{\mathcal{P}\llceil\mathcal{RS}\rrfloor}  =
  \delta C_{\mu\nu\rho}{}^{\mathcal{P}\llceil\mathcal{RS}\rrfloor} - 3\,
  \delta A_{[\mu}{}^{\llceil\mathcal{P}}
  \,B_{\nu\rho]}{}^{\mathcal{RS}\rrfloor}  
  - 2\, A_{[\mu}{}^{\llceil\mathcal{P}} A_\nu{}^{\llceil\mathcal{R}}\delta
  A_{\rho]}{}^{\mathcal{S}\rrfloor\rrfloor}  \,.
\end{equation}
Now we consider again a combined gauge transformation. Requiring that
$\mathcal{H}_{\mu\nu\rho}{}^{\mathcal{MN}}$ transforms covariantly,
it follows that we must choose 
\begin{equation}
  \label{eq:def-gauge-C1}
  \Delta
    C_{\mu\nu\rho}{}^{\mathcal{P}\llceil\mathcal{RS}\rrfloor}
  \Big\vert_\mathrm{gauge} =
  3\,D_{[\mu}\Phi_{\nu\rho]}{}^{\mathcal{P}\llceil\mathcal{RS}\rrfloor}
    + 3\, \mathcal{H}_{\mu\nu}{}^{\llceil\mathcal{P}}
    \,\Xi_\rho{}^{\mathcal{RS}\rrfloor}  
   + \Lambda^{\llceil\mathcal{P}}
    \mathcal{H}_{\mu\nu\rho}{}^{\mathcal{RS}\rrfloor} + 
    \cdots \,, 
\end{equation}
where the unspecified contributions vanish upon contracting $\Delta
C_{\mu\nu\rho}{}^{\mathcal{P}\llceil\mathcal{RS}\rrfloor}$ with
$Y^\mathcal{MN}{}_{\mathcal{P}\llceil\mathcal{RS}\rrfloor}$, so that
they remain as yet undetermined. Here we made use of the Bianchi
identity,
\begin{equation}
  \label{eq:Bianchi-H2}
  D_{[\mu}\mathcal{H}_{\nu\rho]}{}^{\mathcal{M}} = \ft13
  g\,Z^\mathcal{M}{}_\mathcal{NP} \,
  \mathcal{H}_{\mu\nu\rho}{}^{\mathcal{NP}} \,. 
\end{equation}
Note that the standard Bianchi is obtained upon contraction with the
embedding tensor.

At this point we must verify that the algebra of the various
gauge transformations defined so far, will close under
commutation. Let us first summarize the various transformation rules,
\begin{eqnarray}
  \label{eq:gauge-tr-ABC}
  \delta A_\mu{}^\mathcal{M} &=&  D_\mu\Lambda^\mathcal{M} -
  g\,Z^\mathcal{M}{}_{\mathcal{NP}}\,\Xi_\mu{}^{\mathcal{NP}}\,,
  \nonumber\\
  \delta B_{\mu\nu}{}^{\mathcal{MN}} 
  &=& 
    2\,D_{[\mu}\Xi_{\nu]}{}^{\mathcal{MN}} -2\, 
  \Lambda^{\llceil\mathcal{M}}\mathcal{H}_{\mu\nu}{}^{\mathcal{N}\rrfloor}
  + 2\, A_{[\mu}{}^{\llceil\mathcal{M}}
  \delta A_{\nu]}{}^{\mathcal{N}\rrfloor}  \nonumber\\
  &&{}
   - g\, Y^\mathcal{MN}{}_{\mathcal{P}\llceil\mathcal{RS}\rrfloor}
   \, \Phi_{\mu\nu}{}^{\mathcal{P}\llceil\mathcal{RS}\rrfloor} \,,
   \nonumber \\
   \delta
    C_{\mu\nu\rho}{}^{\mathcal{P}\llceil\mathcal{RS}\rrfloor} 
    &=&
    3\,D_{[\mu}\Phi_{\nu\rho]}{}^{\mathcal{P}\llceil\mathcal{RS}\rrfloor}
    + 3\, \mathcal{H}_{\mu\nu}{}^{\llceil\mathcal{P}}
    \,\Xi_\rho{}^{\mathcal{RS}\rrfloor}  
   + \Lambda^{\llceil\mathcal{P}}
    \mathcal{H}_{\mu\nu\rho}{}^{\mathcal{RS}\rrfloor}  
    + 3\,\delta A_{[\mu}{}^{\llceil\mathcal{P}}
    \,B_{\nu\rho]}{}^{\mathcal{RS}\rrfloor}  \nonumber\\ 
    &&{}
    + 2\, A_{[\mu}{}^{\llceil\mathcal{P}} A_\nu{}^{\llceil\mathcal{R}}\delta
    A_{\rho]}{}^{\mathcal{S}\rrfloor\rrfloor} + \cdots \,. 
\end{eqnarray}
These tranformations indeed yield a closed algebra, 
\begin{eqnarray}
  \label{eq:gauge-commutators2}
  {}[\delta(\Lambda_1),\delta(\Lambda_2)] &=& \delta(\Lambda_3) +
  \delta(\Xi_3) + \delta(\Phi_3) \,,  \nonumber \\
  {}[\delta(\Lambda),\delta(\Xi)] &=& \delta(\Phi_4) \,,
  \nonumber\\ 
  {}[\delta(\Xi_1),\delta(\Xi_2)] &=& \delta(\Phi_5)\,,
  \nonumber\\
  {}[\delta(\Lambda),\delta(\Phi)] &=&\cdots \,,
  \nonumber\\ 
  {}[\delta(\Xi),\delta(\Phi)] &=&\cdots \,,
  \nonumber\\ 
  {}[\delta(\Phi_1),\delta(\Phi_2)] &=& 0 \,,
\end{eqnarray}
where we will comment on the two unspecified commutators in a sequal. The
transformation parameters appearing on the right-hand side of
(\ref{eq:gauge-commutators2}) take the following form, 
\begin{eqnarray}
  \label{eq:gauge-parameters2}
  \Lambda_3{}^\mathcal{M} &=& g\,X_{[\mathcal{NP}]}{}^\mathcal{M} \,
  \Lambda_2^\mathcal{N}\Lambda_1^\mathcal{P}\,, \nonumber\\
  \Xi_{3 \mu}{}^{\mathcal{MN}} &=&  \Lambda_1^{\llceil\mathcal{M}}
  D_\mu\Lambda_2^{\mathcal{N}\rrfloor} -
  \Lambda_2^{\llceil\mathcal{M}} D_\mu \Lambda_1^{\mathcal{N}\rrfloor} 
  \,, \nonumber\\
  \Phi_{3 \mu\nu}{}^{\mathcal{P}\llceil\mathcal{MN}\rrfloor}  &=&{}
  \mathcal{H}_{\mu\nu}{}^{\llceil\llceil\mathcal{M}} \left(
  \Lambda_2{}^{\mathcal{N}\rrfloor}\Lambda_1{}^{\mathcal{P}\rrfloor}
   - \Lambda_1{}^{\mathcal{N}\rrfloor}\Lambda_2{}^{\mathcal{P}\rrfloor}
  \right) \,, 
  \nonumber\\
 \Phi_{4 \mu\nu}{}^{\mathcal{P}\llceil\mathcal{MN}\rrfloor}  &=&{}
  2\, D_{[\mu}\Lambda^{\llceil\mathcal{P}}
  \Xi_{\nu]}{}^{\mathcal{MN}\rrfloor} \,,
  \nonumber\\
  \Phi_{5 \mu\nu}{}^{\mathcal{P}\llceil\mathcal{MN}\rrfloor}  &=&{} 
   -g Z^{\llceil\mathcal{P}}{}_\mathcal{RS}
  \left(\Xi_{1[\mu}{}^{\mathcal{MN}\rrfloor} 
  \Xi_{2\nu]}{}^\mathcal{RS}  - \Xi_{2[\mu}{}^{\mathcal{MN}\rrfloor}
  \Xi_{1\nu]}{}^{\mathcal{RS}} \right)\,,
\end{eqnarray}
where the first two equations were already given in
(\ref{eq:gauge-parameters}). 

Continuing this pattern one can derive the full hierarchy of $p$-forms
by iteration. For instance, the transformation rule for
$C_{\mu\nu\rho}{}^{\mathcal{P}\llceil\mathcal{RS}\rrfloor}$ contains
the expected gauge transformation
\begin{equation}
  \label{eq:C-gauge}
  \delta
  C_{\mu\nu\rho}{}^{\mathcal{P}\llceil\mathcal{RS}\rrfloor}= -g
  \Lambda^\mathcal{Q} X_{\mathcal{Q}\llceil\mathcal{K}
  \llceil\mathcal{LM}\rrfloor\rrfloor} 
  {}^{{\llceil\mathcal{P}\llceil\mathcal{RS}\rrfloor\rrfloor}} \;
    C_{\mu\nu\rho}{}^{\mathcal{K}\llceil\mathcal{LM}\rrfloor}\,,
\end{equation}
(where again, $X_{\mathcal{Q}\llceil\mathcal{K}
  \llceil\mathcal{LM}\rrfloor\rrfloor}
{}^{{\llceil\mathcal{P}\llceil\mathcal{RS}\rrfloor\rrfloor}} =
(X_\mathcal{Q})_{\llceil\mathcal{K}
  \llceil\mathcal{LM}\rrfloor\rrfloor}
{}^{{\llceil\mathcal{P}\llceil\mathcal{RS}\rrfloor\rrfloor}}$) up to a
term
\begin{equation}
  \label{eq:delta-C-4}
  \delta 
  C_{\mu\nu\rho}{}^{\mathcal{P}\llceil\mathcal{RS}\rrfloor}= -g\,
  Y^{\mathcal{P}\llceil\mathcal{RS}\rrfloor}{}_{\mathcal{Q}\llceil\mathcal{P}
  \llceil\mathcal{RS}\rrfloor\rrfloor} \,
  \Phi_{\mu\nu\rho}{}^{\mathcal{Q}\llceil\mathcal{P}
  \llceil\mathcal{RS}\rrfloor\rrfloor} \;,
\end{equation}
which characterizes a new gauge transformation with parameter
$\Phi_{\mu\nu\rho}
{}^{\mathcal{Q}\llceil\mathcal{P}\llceil\mathcal{RS}\rrfloor\rrfloor}$,
associated with a new four-rank tensor field which will again belong
to some restricted subrepresentation. It turns out that the two
unspecified commutators in (\ref{eq:gauge-commutators2}) are precisely
given by these transformations. The tensor
$Y^{\mathcal{P}\llceil\mathcal{RS}\rrfloor}{}_{\mathcal{Q}\llceil\mathcal{P}
  \llceil\mathcal{RS}\rrfloor\rrfloor}$ acts as an intertwiner between
the three- and four-rank tensor fields, and can easily be written down
explicitly,
\begin{eqnarray}
  \label{eq:higherY-p=3}
  Y^{{\cal K}\llceil{\cal MN}\rrfloor}{}_{{\cal P}\llceil{\cal Q}\llceil
  \mathcal{RS}\rrfloor\rrfloor}   
   &=&{} 
   - \delta^{\llceil {\cal K}}_{\mathcal{P}}\;
   Y_{\vphantom{\llceil}}^{{\cal MN}\rrfloor}
  {}^{\vphantom{\llceil}}_{{\cal Q}\llceil\mathcal{RS}\rrfloor}  
    -X_{{\cal P}\llceil {\cal Q}\llceil {\cal RS}\rrfloor\rrfloor}
   {}^{\llceil\mathcal{K}\llceil\mathcal{MN}\rrfloor\rrfloor}
  \nonumber\\[1ex]
  &=&{} -  2\,\Big(
  \delta_{\cal Q}^{\cal\llceil  K}\delta_{\cal S}^{\cal \llceil
    M}\,X_{\cal PR}{}_{\vphantom{\llceil}}^{\cal N\rrfloor\rrfloor} 
  +\delta_{\cal P}^{\cal\llceil  K}\delta_{\cal Q}^{\cal\llceil
    M}\,X_{\cal SR}{}_{\vphantom{\llceil}}^{\cal N\rrfloor\rrfloor} 
  \Big)  \nonumber\\[.5ex]
  &&{}
  +2\,\Big( \delta_{\cal P}^{\cal\llceil  K}\delta_{\cal S}^{\cal\llceil
    M}\,X_{\cal QR}{}_{\vphantom{\llceil}}^{\cal N\rrfloor\rrfloor} 
  +\delta_{\cal R}^{\cal\llceil  K}\delta_{\cal S}^{\cal\llceil
    M}\,X_{\cal PQ}{}_{\vphantom{\llceil}}^{\cal N\rrfloor\rrfloor} 
  \Big)    \;.
\end{eqnarray}
To derive the second formula we made use of (\ref{eq:Y-symmetrized}).
Observe that on the right-hand side we must apply the projector
(\ref{eq:Y-proj}) in order to obtain the restricted representations in
the index triples
$\llceil\mathcal{K}\llceil\mathcal{MN}\rrfloor\rrfloor$ and
$\llceil\mathcal{Q}\llceil\mathcal{RS}\rrfloor\rrfloor$, respectively;
the result is then automatically projected onto a restricted
representation in the indices
$\mathcal{P}\llceil\mathcal{Q}\llceil\mathcal{RS}\rrfloor\rrfloor$.

At this point one recognizes that there exists a whole
hierarchy of such tensors.\footnote{%%%%%%%%%%%%%%%%%%%%%%%%%%
  From this point we denote the intertwining tensors and $p$-forms by
  $Y$ and $C$, respectively, and the corresponding gauge
  transformation parameters by $\Phi$. Their rank will be obvious from
  the index structure.  } %%%%%%%%%%%%%%%%%%%%%%%%%%%%%%%%%%%% 
They are defined by ($p\geq3$)
\begin{eqnarray}
  \label{eq:higherY}
  Y^{{\cal M}_1\llceil{\cal M}_2\llceil
  \cdots\mathcal{M}_p\rrfloor\cdot\cdot\rrfloor}  
     {}_{{\cal N}_0\llceil{\cal N}_1\llceil
  \cdots\mathcal{N}_p\rrfloor\cdot\cdot\rrfloor}   
   &=&{} 
   - \delta^{\llceil {\cal M}_1}_{{\cal N}_0}\;
   Y_{\vphantom{\llceil}}^{{\cal M}_2\llceil
  \cdots\mathcal{M}_p\rrfloor\cdot\cdot\rrfloor}
  {}^{\vphantom{\llceil}}_{{\cal N}_1\llceil 
  {\cal N}_2\llceil\cdots\,\mathcal{N}_p\rrfloor\cdot\cdot\rrfloor} 
   \nonumber\\[1.5ex]
    &&{}
    -X_{{\cal N}_0\llceil {\cal N}_1\llceil {\cal N}_2\llceil\cdots\, 
      \mathcal{N}_p\rrfloor\cdot\cdot\rrfloor}
   {}^{\llceil\mathcal{M}_1\llceil\mathcal{M}_2\llceil\cdots\,  
  \mathcal{M}_p\rrfloor\cdot\cdot\rrfloor }\;\;\;, 
\end{eqnarray}
where, as before, we employ the notation,
\begin{equation}
  \label{eq:1X-(X)}
  X_{{\cal N}_0\llceil {\cal N}_1\llceil {\cal N}_2
    \llceil\cdots\, \mathcal{N}_p\rrfloor\cdot\cdot\rrfloor}
  {}^{\llceil\mathcal{M}_1\llceil\mathcal{M}_2\llceil\cdots\,
    \mathcal{M}_p\rrfloor\cdot\cdot\rrfloor}=
  (X_{\mathcal{N}_0})_{\llceil {\cal N}_1\llceil {\cal N}_2\llceil\cdots\,
    \mathcal{N}_p\rrfloor\cdot\cdot\rrfloor}
  {}^{\llceil\mathcal{M}_1\llceil\mathcal{M}_2\llceil\cdots\,
    \mathcal{M}_p\rrfloor\cdot\cdot\rrfloor}\;\;.
\end{equation}
All these tensors are gauge invariant and they are formed from the
embedding tensor multiplied by invariant tensors of the duality group
$\mathrm{G}$, so that they all transform in (a subset of) the same
representations as the embedding tensor. By induction, one can prove
their mutual orthogonality,
\begin{eqnarray}
  \label{eq:orthoY}
  Y^{{\cal K}_2\llceil{\cal K}_3\llceil\cdots\mathcal{K}_p\rrfloor
  \cdot\cdot\rrfloor}
  {}_{{\cal M}_1\llceil\mathcal{M}_2\llceil\cdots
  \mathcal{M}_p\rrfloor\cdot\cdot\rrfloor} \; 
   Y^{{\cal M}_1\llceil\mathcal{M}_2\llceil\cdots
  \mathcal{M}_p\rrfloor\cdot\cdot\rrfloor}
  {}_{{\cal N}_0\llceil{\cal N}_1\llceil\cdots
  \mathcal{N}_p\rrfloor\cdot\cdot\rrfloor}  
   &=&{} 0 \;.
\end{eqnarray}
To see this, one substitutes the expression (\ref{eq:higherY}) for the
second $Y$-tensor and uses the gauge invariance of the first $Y$-tensor
to obtain the expression, 
\begin{eqnarray}
  \label{eq:orthoY-proof}
  \mathrm(\ref{eq:orthoY})
  &=&{}
  - Y^{{\cal K}_2\llceil{\cal K}_3\llceil\cdots\mathcal{K}_p\rrfloor
  \cdot\cdot\rrfloor}
  {}_{\mathcal{N}_0\llceil\mathcal{M}_2\llceil\cdots
  \mathcal{M}_p\rrfloor\cdot\cdot\rrfloor} \;\, 
   Y^{\mathcal{M}_2\llceil\mathcal{M}_3\llceil\cdots
  \mathcal{M}_p\rrfloor\cdot\cdot\rrfloor}
  {}_{{\cal N}_1\llceil{\cal N}_2\llceil\cdots
  \mathcal{N}_p\rrfloor\cdot\cdot\rrfloor}  \nonumber\\[1.5ex]
   &&{} 
   - Y^{{\cal M}_2\llceil{\cal M}_3\llceil\cdots\mathcal{M}_p\rrfloor
  \cdot\cdot\rrfloor}
  {}_{\mathcal{N}_1\llceil\mathcal{N}_2\llceil\cdots
  \mathcal{N}_p\rrfloor\cdot\cdot\rrfloor} \;\, 
    X_{\mathcal{N}_0\llceil\mathcal{M}_2\llceil\mathcal{M}_3\llceil\cdots
  \mathcal{M}_p\rrfloor\cdot\cdot\rrfloor}
  {}^{\llceil{\cal K}_2\llceil\mathcal{K}_3\llceil\cdots
  \mathcal{K}_p\rrfloor\cdot\cdot\rrfloor} 
 \;.
\end{eqnarray}
This result vanishes upon expressing the generator $X$ on the
right-hand side in terms of the $Y$-tensors, using the definition
(\ref{eq:higherY}), and subsequently using the orthogonality
constraint for a lower value of the rank $p$. The fact that
symmetrization over the three last indices of the restricted
representation will vanish as a result of (\ref{eq:Y-symmetrized}),
implies that higher-rank tensors will vanish as well under certain
index symmetrizations.

The $Y$-tensors form an (infinite, in principle) hierarchy of
intertwiners between successive sets of restricted representations of
tensor gauge fields. The restrictions on the representations occurring
at the $(p\!+\! 1)$-th step of the iteration are determined
inductively via formula (\ref{eq:higherY}), where on the right-hand
side the projectors obtained at the previous $p$-th step of the
iteration must be applied to the $p$-tuples of indices ${\mathcal
  M}_1\llceil
\mathcal{M}_2\llceil\cdots\mathcal{M}_p\rrfloor\cdot\cdot\rrfloor$ and
${\mathcal N}_1\llceil\mathcal{N}_2\llceil\cdots \mathcal{N}_p\rrfloor
\cdot\cdot\rrfloor$, respectively. We emphasize that {\em no other
  information is needed to determine the hierarchy}. However, as we
pointed out already, working out more explicit expressions requires a
case-by-case study, as we will exemplify for $D=3$ maximal
supergravity with duality group $\mathrm{G} = \mathrm{E}_{8(8)}$ in
section~\ref{sec:three-dimensions} of this paper. Consequently, given
the $Y$-tensors, and specifying the duality group G, the above results
enable a complete determination of the full hierarchy of the
higher-rank $p$-forms required for the consistency of the gauging. In
particular, we can exhibit some of the terms in the variations of the
$p$-form fields that follow rather directly from the previous
discussion,
\begin{eqnarray}
  \label{eq:trans-C-p}
  \delta C_{\mu_1\dots\mu_p}{}^{\mathcal{M}_1\llceil
  \mathcal{M}_2\llceil\cdots\mathcal{M}_p\rrfloor\cdot\cdot\rrfloor}
  &=&{}
  p\, D_{[\mu_1}\Phi_{\mu_2\cdots\mu_p]}{}^{\mathcal{M}_1\llceil
  \mathcal{M}_2\llceil\cdots\mathcal{M}_p\rrfloor\cdot\cdot\rrfloor} 
   \nonumber \\[1.5ex]
   &&{}
   + \Lambda^{\llceil\mathcal{M}_1}
    \mathcal{H}_{\mu_1\cdots\mu_p}{}^{\llceil\mathcal{M}_2
  \cdots\rrfloor\cdot\cdot\rrfloor\rrfloor} 
   +p\,\delta A_{[\mu_1}{}^{\llceil\mathcal{M}_1}
    \,C_{\mu_2\cdots\mu_p]}{}^{\llceil\mathcal{M}_2
  \cdots\rrfloor\cdot\cdot\rrfloor\rrfloor}  
    \nonumber\\[1.5ex]
   &&{} 
   -g\,Y^{{\cal M}_1\llceil{\cal M}_2\llceil\cdots
  \mathcal{M}_p\rrfloor\cdot\cdot\rrfloor}  
   {}_{{\cal N}_0\llceil\mathcal{N}_1\llceil
  \dots\mathcal{N}_p\rrfloor.\rrfloor} \; 
   \Phi_{\mu_1\dots\mu_p}{}^{{\cal N}_0\llceil{\cal N}_1\llceil
  \dots\mathcal{N}_p\rrfloor\cdot\cdot\rrfloor}\;,  
   \nonumber\\[1.5ex]
   &&{}+  \cdots\;. 
\end{eqnarray}

Although the number of space-time dimensions does not enter into this 
analysis (as we said, the iteration procedure can in principle be 
continued indefinitely) there is, for the maximal supergravities, a
consistent correlation between the rank of the tensor fields and the
occurrence of conjugate $\mathrm{G}$-representations that is precisely
in accord with tensor-tensor and vector-tensor (Hodge)
duality\footnote{%%%%%%%%%%%%% 
  As well as with the count of physical degrees of freedom.}
corresponding to the space-time dimension where the maximal
supergravity with that particular duality group $\mathrm{G}$ lives. In
the next section we discuss some of the results of this analysis and
their implications for M-theory degrees of freedom.

%%%%%%%%%%%%%%%%%%%%%%%%%%%%%%%%%%%%%%%%%%%%%%%%%%%%%%%%%%%%%%%%%
\section{M-Theory degrees of freedom}
\setcounter{equation}{0}
\label{sec:m-theory-degrees}
%%%%%%%%%%%%%%%%%%%%%%%%%%%%%%%%%%%%%%%%%%%%%%%%%%%%%%%%%%%%%%%%%
The hierarchy of vector and tensor gauge fields that we presented in
the previous section can be considered in the context of the maximal
gauged supergravities. In that case the gauge group is embedded in the
duality group $\mathrm{G}$, which depends on the space-time dimension
in which the supergravity is defined. Once we specify the group
$\mathrm{G}$ the representations can be determined of the various
$p$-form potentials. In principle the hierarchy allows a unique
determination of the higher $p$-forms, but in practice this
determination tends to be somewhat subtle. To see this, let us first
briefly consider the possible representations for the two-forms. For
that we need the representations in the symmetric product of two
representations belonging to the vector fields (we will deal with the
case $D=3$ separately),
\begin{equation}
  \label{eq:2-forms}
\begin{array}{lcrcl}
  D=7 &:&\overline{\mathbf{10}}\times_\mathrm{sym}\overline{\mathbf{10}} 
  &=& \mathbf{5}+ \overline{\mathbf{50}}\;,
  \nonumber\\ 
  D=6 &:&\mathbf{16}_c \times_\mathrm{sym}\mathbf{16}_c 
  &=& \mathbf{10}+\mathbf{126}_c  \;,
  \nonumber\\ 
  D=5 &:&\overline{\mathbf{27}}\times_\mathrm{sym}\overline{\mathbf{27}} 
  &=& \mathbf{27}+ \overline{\mathbf{351}}^{\,\prime}  \;,
  \nonumber \\ 
  D=4 &:&\mathbf{56} \times_\mathrm{sym} \mathbf{56} 
  &=& \mathbf{133}+ \mathbf{1463}  \;. 
%  D=3 &&\mathbf{248} \times_\mathrm{sym} \mathbf{248} 
%  &=& \mathbf{}\mathbf{}\mathbf{}\mathbf{}
\end{array} 
\end{equation}
Hence it seems that the two-forms can belong to two possible
representations of the duality group. To see which representation is
allowed, we take its conjugate and consider once more the product with
the vector field representation, This product should contain the
representation associated with the tensor
$Z^\mathcal{M}{}_\mathcal{NP}$. The latter is simply equal to the
representation of the embedding tensor. If this representation is
contained in the product, then we are dealing with an acceptable
candidate representation. If this is not the case, then we must
conclude that $Z^\mathcal{M}{}_\mathcal{NP}$ cannot act as an
intertwiner between the corresponding two-forms and the one-form
potentials.

Performing this test\footnote{%%%%%%%%%%%%%%%%%%%%%%%%%%%%%%%%%
  We used the Lie package \cite{LeCoLi92} for computing such
  decompositions. } % 
%%%%%%%%%%%%%%%%%%%%%%%%%%%%%%%%%%%%%%%%%%%%%%%%%%%%%%%%%%%%%%%%
on each of the two representations in (\ref{eq:2-forms}), 
it turns out that only the first representation is allowed, leading to
the entries for the two-forms presented in the third column of table
\ref{tab:vector-tensor-repr}. For the case of $D=3$ space-time
dimensions the above approach leads only to a partial determination of
the representation assignment. Here the symmetric product decomposes
into six different representations and in section
\ref{sec:three-dimensions} we will proceed diffently to deduce the
correct assignment. The results for the two-forms in $4\leq D\leq 7$
dimensions were originally derived in \cite{deWit:2005hv}, where also
the representations of the three-forms were determined that are shown
in the table.

As we stressed already the hierarchy leads to a unique determination
of the representations of the higher-rank tensor fields, but this has
only partially been carried out. Already for lower-rank tensors, table
\ref{tab:vector-tensor-repr} shows remarkable features.  We recall
that the analysis described in section~\ref{sec:hierarchy} did not
depend on the number of space-time dimensions. For instance, it is
possible to derive representation assignments for $(D\!+\!1)$-rank
tensors, although these do not live in a $D$-dimensional space-time.
On the other hand, whenever there exists a (Hodge) duality relation
between fields of different rank at the appropriate value for $D$,
then one finds that their $\mathrm{G}$ representations turn out to be
related by conjugation. This property is already exhibited at the
level of the lower-rank tensors and we have simply extrapolated this
pattern to higher-rank fields. Furthermore the diagonals pertaining to
the $(D\!-\!2)$-, $(D\!-\!1)$- and $D$-rank tensor fields refer to the
adjoint representation and the representations conjugate to those
assigned to the embedding tensor and its quadratic constraint,
respectively. While not all of these features show up fully for the
lower-rank tensors, the pattern is quite suggestive. The underlying
reasons for some of this will become apparent in the later sections,
where we establish that the $(D\!-\!1)$- and $D$-rank tensors play the
special role of imposing the constancy of the embedding tensor and the
closure of the corresponding gauge group.

%%%%%%%%%%%%%%%%%%%%%%%%%%%%%%%%%%%%%%%%%%%%%%%%%%%%%%%%%%%%%%%%%%
%%%%%%%%%%%%%%%%%%%%%%%%%%%%%%%%%%%%%%%%%%%%%%%%%%%%%%%%%%%%%%%%%%
\begin{table}[t]
\centering
\begin{tabular}{l l cccccc  }\hline
~&~&~&~&~&~&~& \\[-4mm]
~ &$~$& 1&2&3&4&5&6  \\   \hline
~&~&~&~\\[-4mm]
7   & ${\rm SL}(5)$ & $\overline{\bf 10}$  & ${\bf 5}$ & $\overline{\bf 5}$ &
${\bf 10}$ &  ${\bf 24}$ & $\overline{\bf 15}+{\bf 40}$  \\[1mm]
6  & ${\rm SO}(5,5)$ & ${\bf 16}_c$ & ${\bf 10}$ & ${\bf 16}_s$ & 
  ${\bf 45}$ & ${\bf 144}_s$ &  
$\!\!{\bf 10}\!+\! {\bf 126}_s\!+\! {\bf 320}\!\!$\\[.8mm]
5   & ${\rm E}_{6(6)}$ & $\overline{\bf 27}$ & ${\bf 27}$ & ${\bf 78}$
& ${\bf 351}$ & $\!\!{\bf 27}\!+\! {\bf 1728}\!\!$ &  \\[.5mm]
4   & ${\rm E}_{7(7)}$ & ${\bf 56}$ & ${\bf 133}$ & ${\bf 912}$ &
$\!\!\!{\bf 133}\!+\! {\bf 8645}\!\!\!$ &    \\[.5mm]
3   & ${\rm E}_{8(8)}$ & ${\bf 248}$ & ${\bf 1}\!+\! {\bf 3875}$ & ${\bf
  3875}\!+\! {\bf147250}$ & & 
\\ \hline
\end{tabular}
%%%%%%%%%%%%%%%%%%%%%%%%%%%%%%%%%%%%%%%%%%%%%%%%%%%%%%%%%%%%%%%%%
\caption{\small
Duality representations of the vector and tensor gauge fields
for gauged  maximal supergravities in space-time dimensions $3\leq
D\leq 7$. The first two columns list the space-time dimension and the
corresponding duality group. Note that the singlet two-form in three
dimensions is not induced by the hierarchy. Its presence follows from
independent considerations, which are discussed in the text. 
}\label{tab:vector-tensor-repr}
\end{table}
%%%%%%%%%%%%%%%%%%%%%%%%%%%%%%%%%%%%%%%%%%%%%%%%%%%%%%%%%%%%%%%%%%
%%%%%%%%%%%%%%%%%%%%%%%%%%%%%%%%%%%%%%%%%%%%%%%%%%%%%%%%%%%%%%%%%%

It is an obvious question whether these systematic features have a
natural explanation in terms of M-theory. Supergravity may contain
some of the fields carrying charges that could induce a gauging. For
instance, in the toroidal compactification there are towers of massive
Kaluza-Klein states whose charges couple to the corresponding
Kaluza-Klein gauge fields emerging from the higher-dimensional metric.
This is of direct relevance in the so-called Scherk-Schwarz reductions
\cite{Scherk:1979zr}.  However, these Kaluza-Klein states cannot
generally be assigned to representations of the duality group and
therefore there must be extra degrees of freedom whose origin cannot
be understood within the context of a dimensional compactification of
supergravity.\footnote{%%%%%%%%%%%%%%%%%%%%%%%%%%%%%%%%%%%%%%%%%%
  In view of the fact that the Kaluza-Klein states are 1/2-BPS, also
  these extra degrees of freedom must correspond to 1/2-BPS states. %
} %%%%%%%%%%%%%%%%%%%%%%%%%%%%%%%%%%%%%%%%%%%%%%%%%% 
This phenomenon was discussed some time ago, for
instance, in \cite{Obers:1998fb,deWit:2000wu}. 

The general gaugings that have been constructed in recent years 
obviously extend beyond gaugings whose charges are carried by
supergravity degrees of freedom. The embedding tensor can be regarded
as a duality covariant tensor that, once it is fixed to some constant
value, selects a certain subsector of the available charge
configurations carried by degrees of freedom that will cover complete
representations of the duality group. If this idea is correct these
degrees of freedom must exist in M-theory, and there are indeed
indications that this is the case. In this way the gauging acts
as a probe of M-theory degrees of freedom.

Independent evidence that this relation with M-theory degrees of freedoms
is indeed realized is provided by the work of \cite{Elitzur:1997zn}
(see also, \cite{Obers:1998fb} and references quoted therein) where
matrix theory \cite{deWit:1988ig,Banks:1996vh} is considered in a toroidal
compactification.  These results are based on the correspondence
between $N=4$ super-Yang-Mills theory on a (rectangular) spatial torus
$\tilde{T}^n$ with radii $s_1,\ldots, s_n$, and M-theory in the
infinite-momentum frame on the dual torus $T^n$ with radii $R_1,R_2,
\ldots, R_n$, where $s_i= l_\mathrm{p}^3/R_{11}R_i$ and $l_\mathrm{p}$
denotes the Planck length in eleven dimensions. The conjecture then is
that the latter should be invariant under permutations of the radii
$R_i$ and under T-duality of type-IIA string theory. The relevant
T-duality transformations follow from making two consecutive
T-dualities on two different circles. When combined with the
permutation symmetry, T-duality can be represented by
($i\not=j\not=k\not=i$)
\begin{eqnarray}
  \label{eq:T-duality}
  R_i\to \frac{l_\mathrm{p}^3}{R_j R_k} \;, \quad 
  R_j\to \frac{l_\mathrm{p}^3}{R_k R_i} \;, \quad 
  R_k\to \frac{l_\mathrm{p}^3}{R_i R_j} \;, \quad 
  l_\mathrm{p}^3 \to \frac{l_\mathrm{p}^6}{R_i R_j R_k} \;,
\end{eqnarray}

The above transformations generate a discrete group which coincides
with the Weyl group of $\mathrm{E}_n$; on the Yang-Mills side, the
elementary Weyl reflections correspond to permutations of the
compactified coordinates (generating the Weyl group of
$\mathrm{SL}(n)$) and Montonen-Olive duality
$g_\mathrm{eff}\rightarrow 1/g_\mathrm{eff}$ (corresponding to
reflections with respect to the exceptional node of the $\mathrm{E}_n$
Dynkin diagram).  This Weyl group, which leaves the rectangular shape
of the compactification torus invariant, can be realized as a discrete
subgroup of the compact subgroup of $\mathrm{E}_{n(n)}$, and
consequently as a subgroup of the conjectured non-perturbative duality
group $\mathrm{E}_{n(n)}(\mathbb{Z})$ \cite{Hull:1994ys}.
Representations of this symmetry can now be generated by mapping out
the Weyl orbits starting from certain states. For instance, one may
start with Kaluza-Klein states on $T^n$, whose masses are proportional
to $M\sim 1/R_i$. The action of the Weyl group then generates new
states, such as the ones that can be identified with two-branes
wrapped around the torus, whose masses are of order $M\sim
R_jR_k/l_\mathrm{p}^3$, and so on.  According to \cite{Hull:1994ys},
the non-perturbative states should combine into multiplets of
$\mathrm{E}_{n(n)}(\mathbb{Z})$; if the representation has weights of
different lengths, one needs several different Weyl orbits to recover
all states in the representation.

Following this procedure one obtains complete multiplets of the
duality group (taking into account that some states belonging to the
representation will vanish under the Weyl group and will therefore
remain inaccessible by this construction). More specifically, using
the relation $n=11-D$, it turns out that the first two columns of
table~\ref{tab:vector-tensor-repr}, respectively, correspond to the
so-called flux and momentum multiplets of \cite{Elitzur:1997zn}.
However, as already pointed out above, the conjecture of
\cite{Hull:1994ys} is essential in that one may need extra states from
different Weyl orbits in order to get the full representation; for
instance, there are only 2160 momentum states for $\mathrm{E}_{8(8)}$,
which must be supplemented by 8-brane states to obtain the full
$\bf{3875}$ representation of $\mathrm{E}_{8(8)}$.

The representations in the table were also found in
\cite{Iqbal:2001ye}, where a `mysterious duality' was exhibited
between toroidal compactifications of M-theory and del Pezzo surfaces.
Here the M-theory dualities are related to global diffeomorphisms that
preserve the canonical class of the del Pezzo surface. Again the
representations thus found are in good agreement with the
representations in table~\ref{tab:vector-tensor-repr}.

For $n\geq 9$, the flux and momentum multiplets of
\cite{Elitzur:1997zn} have infinitely many components. Indeed, there
are hints that the above considerations concerning new M-theoretic
degrees of freedom can be extended to infinite-dimensional duality
groups. Already some time ago \cite{West:2004kb} it was shown from an
analysis of the indefinite Kac--Moody algebra $\mathrm{E}_{11}$ that
the decomposition of its so-called L1 representation at low levels
under its finite-dimensional subalgebra $\mathrm{SL}(3) \times
\mathrm{E}_{8}$ yields the same $\mathbf{3875}$ representation that
appears for the two-forms as shown in
table~\ref{tab:vector-tensor-repr}. This analysis has meanwhile been
extended \cite{Riccioni:2007au,Bergshoeff:2007qi,Bergshoeff:2007vb} to
other space-time dimensions and higher-rank forms, and again there is
a clear overlap with the representations in
table~\ref{tab:vector-tensor-repr}.  Nevertheless it remains far from
clear what all these (infinitely many) new degrees of freedom would
correspond to, and how they would be concretely realized. Concerning
the physical interpretation of the new states, a first step was taken
in \cite{Englert:2007qb}, where an infinite multiplet of BPS states is
generated from the M2 brane and M5 brane solutions of $D=11$
supergravity by the iterated action of certain $A_1^{(1)}$ subgroups
of the $\mathrm{E}_9$ Weyl group. In the context of gauged
supergravities, the significance of these states may become clearer
with the exploration of maximal gauged supergravities in {\it two}
space-time dimensions \cite{SamtlebenWeidner2}, where the embedding
tensor transforms in the so-called basic representation of
$\mathrm{E}_9$ (which is infinite dimensional).

%%%%%%%%%%%%%%%%%%%%%%%%%%%%%%%%%%%%%%%%%%%%%%%%%%%%%%%%%%%%%%%%%%
%%%%%%%%%%%%%%%%%%%%%%%%%%%%%%%%%%%%%%%%%%%%%%%%%%%%%%%%%%%%%%%%%%
\section{Tensor field representations in three space-time
  dimensions } 
\setcounter{equation}{0}
\label{sec:three-dimensions}
%%%%%%%%%%%%%%%%%%%%%%%%%%%%%%%%%%%%%%%%%%%%%%%%%%%%%%%%%%%%%%%%%%
%%%%%%%%%%%%%%%%%%%%%%%%%%%%%%%%%%%%%%%%%%%%%%%%%%%%%%%%%%%%%%%%%%
Here and in the following two sections we will illustrate the
preceding discussion and consider maximal supergravity in three
space-time dimensions, where the full tensor hierarchy of $p$-forms is
short enough to obtain all relevant information from the explicit
results given in section~\ref{sec:hierarchy}.  This example will show
all the characteristic features that are generic for gauged
supergravities. In this section we will determine the representation
assignments for the tensor fields. The relevant duality group is equal
to $\mathrm{E}_{{8(8)}}$, which is of dimension 248. Its fundamental
representation coincides with the adjoint representation, so that the
generators in this representation are given by the ${\rm E}_{{8(8)}}$
structure constants, $(t_\mathcal{M})_\mathcal{N}{}^{\mathcal{P}} =
-f_{\mathcal{MN}}{}^{\mathcal{P}}$. Indices may be raised and lowered
by means of the Cartan-Killing form $\eta_{{\cal MN}}$.  The vector
fields $A_{\mu}{}^{\mathcal{M}}$ transform in the $\mathbf{248}$
representation and the embedding tensor $\Theta_{{\cal MN}}$ is a {\it
  symmetric} matrix belonging to the $\mathbf{3875}+\mathbf{1}$
representation \cite{Nicolai:2000sc,Nicolai:2001sv}. Using these data,
we may evaluate the general formulas of section~\ref{sec:hierarchy}
for this particular theory.

The gauge group generators are obtained by contracting
$\mathrm{E}_{8(8)}$ generators with the embedding tensor
$X_\mathcal{M}\equiv\Theta_{\cal MN}\,t^{\cal N}$. In the adjoint
representation we thus have
\begin{equation}
  \label{eq:X-MNP-3}
  X_{\mathcal{MN}}{}^{\mathcal{P}}=
  -\Theta_{{\cal M}}{}^\mathcal{Q}\,f_{\mathcal{QN}}{}^{{\cal P}}{}=
  \Theta_{{\cal MQ}}\,f^{{\cal QP}}{}_{{\cal N}} \;.
\end{equation}
The tensor $Z^{\mathcal{P}}{}_{\mathcal{MN}}$ defined in
(\ref{eq:def-Z}) is then given by
\begin{eqnarray}
  \label{eq:Z3}
  Z^{\cal P}{}_{\cal MN}&=&
  \Theta_{{\cal Q(M}}\,f^{{\cal QP}}{}_{{\cal N)}} \;.
\end{eqnarray}
Because this tensor is a group invariant contraction of the embedding
tensor, its representation must overlap with some of the
representations of the embedding tensor. Obviously, the singlet
component drops out so that we may conclude that (\ref{eq:Z3}) must
belong to the $\mathbf{3875}$ representation.

As discussed before ({\it cf.} (\ref{eq:Z-proj})), the tensor
$Z^{\mathcal{K}}{}_{\mathcal{MN}}$ generically does not map onto the
full symmetric tensor product $({\mathcal{MN}})$, which decomposes
according to 
\begin{equation}
  \label{eq:X-adjoint}
  {\bf 248} \times_{\rm sym} {\bf 248} =
  {\bf 1}+
  {\bf 3875}+
  {\bf 27000}\;,
\end{equation} 
but only on a restricted representation. Since (\ref{eq:Z3}) represents an
infinitesimal ${\rm E}_{8(8)}$ transformation on the embedding tensor
$\Theta_{\mathcal{MN}}$ which leaves the representation content
invariant, it follows that the indices $(\mathcal{MN})$ in (\ref{eq:Z3}) are
restricted to the $\mathbf{3875}$  representation, so that the
relevant projector is precisely $\mathbb{P}^{(3875)}$ acting on the
symmetric tensor product.  This projector can be written as \cite{KNS}
\begin{eqnarray}
  \label{eq:proj-3875}
  ({\mathbb P}^{{(3875)}})^{{\cal RS}}{}_{{\cal MN}} &=&
  \ft17\,\delta_{{\cal M}}^{({\cal R}}\delta_{{\cal N}}^{{\cal S})}\,
  -\ft1{56}\,\eta_{{\cal MN}}\,\eta^{{\cal RS}} -\ft1{14}\,f^{\cal
  P}{}_{\cal M}{}^{\cal (R}\,f_{\cal PN}{}^{\cal S)} \;.    
\end{eqnarray}
According to the general discussion, it follows that closure of
the vector field gauge algebra requires the introduction of two-forms
in the ${\bf 3875}$ representation.  Hence the two-forms transform in 
the same representation as the embedding tensor. As noted in the previous
section, this is a general pattern in gauged supergravities: the
embedding tensor in $D$ dimensions 
transforms in the representation which is conjugate to the
$(D\!-\!1)$-forms.  More precisely, the field strength of the
$(D\!-\!1)$-forms is dual to the embedding tensor.  We will discuss
the explicit relation in the next sections.  In three dimensions there
is a subtlety related to the fact that the embedding tensor is not
irreducible but contains an additional singlet ${\bf 1}$ besides the
${\bf 3875}$.  The associated two-form can be defined but does not yet
show up in the tensor hierarchy at this point. In order to keep the
discussion as simple as possible, we will in the following restrict to
the gaugings induced by an embedding tensor in the irreducible~${\bf
  3875}$.

Continuing the tensor hierarchy according to the general pattern
discussed above, the next intertwining tensor $Y^{\cal MN}{}_{\cal
  K\llceil PQ\rrfloor}$, defined in (\ref{eq:def-Y-2}), takes the form
\begin{eqnarray}
  \label{eq:Y2}
  Y^{\cal MN}{}_{\cal K\llceil PQ\rrfloor} &\equiv& 
  2\,\Big(\delta_{\llceil\mathcal{P}}{}_{\vphantom{\llceil}}
  ^{\llceil\mathcal{R}}\, 
  f_{\vphantom{\llceil}}^{\mathcal{S}\rrfloor \llceil 
  \mathcal{M}}{}_{\mathcal{Q}\rrfloor}\, \delta_{\cal
  K}{}_{\vphantom{\llceil}}^{\cal N\rrfloor}    -
  \delta_{\cal K}{}_{\vphantom{\llceil}}^{\llceil \cal R}\,
  f_{\vphantom{\llceil}}^{\cal S\rrfloor \llceil 
  M}{}_{\llceil\mathcal{P}}\,
  \delta_{\mathcal{Q}\rrfloor}{}{}_{\vphantom{\llceil}}^{\mathcal{N}\rrfloor}
  \Big)  \, \Theta_{\cal RS} \;.
\end{eqnarray}
In view of the group-invariant contractions, the tensor
$Y^\mathcal{MN}{}_{\cal K\llceil PQ\rrfloor}$ transforms again in the
$\mathbf{3875}$ representation.  It controls the appearance of
three-forms in the gauge transformations of two-forms and thereby
determines the (minimal) field content of three-forms required for
consistency of the algebra. Again it does not map onto the full tensor
product ${\cal K\llceil PQ\rrfloor}$ but only onto a restricted
subrepresentation, as in~(\ref{eq:Z-proj}). To determine this
subrepresentation, we observe that the expression in parentheses in
(\ref{eq:Y2}) is symmetric under exchange $\llceil
\mathcal{RS}\rrfloor\,\leftrightarrow\,\llceil \mathcal{MN}\rrfloor$, and
thus transforms in
\begin{eqnarray}
  \label{eq:repQ2}
  {\bf 3875} \times_{\rm sym} {\bf 3875}
  &=&{\bf 1}+
  {\bf 3875}+
  {\bf 27000}+
  {\bf 147250}+
  {\bf 2450240}+
  {\bf 4881384}\;.
\end{eqnarray}
On the other hand, by its index structure, the tensor product
$\mathcal{K}\llceil\mathcal{PQ}\rrfloor$ is given by 
\begin{eqnarray}
  \label{eq:repQ1}
  {\bf 248} \times {\bf 3875}
  &=&
  {\bf 248}+
  {\bf 3875}+
  {\bf 30380}+
  {\bf 147250}+
  {\bf 779247}\;,
\end{eqnarray}
Comparing (\ref{eq:repQ2}) and (\ref{eq:repQ1}), it follows that the
index combination $\mathcal{K}\llceil\mathcal{PQ}\rrfloor$ is indeed
restricted to certain irreducible representations so that the three-forms
transform in the representation\footnote{%%%%%%%%%%%%%%%%%%%%%%%%%
  The absence of the ${\bf 248}$, ${\bf 30380}$ and ${\bf 779247}$
  representations is in accord with equation (\ref{eq:Y-symmetrized})
  because those are contained in the fully symmetrized product of three
  $\mathbf{248}$ representations. } %%%%%%%%%%%%%%%%%%%%%%%%%%%%%%
\begin{eqnarray}
  \label{eq:Qrep}
  C_{\mu\nu\rho}{}^{{\cal K\llceil PQ\rrfloor}}&\sim& 
  {\bf 3875}+ {\bf 147250} \;.
\end{eqnarray}
In principle, the argument so far does not exclude the possibility
that the image of $Y^{\cal MN}{}_{\cal K\llceil PQ\rrfloor}$ is
restricted to only one of the two irreducible representations in
(\ref{eq:Qrep}). To show that both irreducible parts are present,
one may {\it e.g.}\ compute and diagonalize the action of the ${\rm
  E}_{8(8)}$ Casimir operator on $Y^{\cal MN}{}_{\cal K\llceil
  PQ\rrfloor}$.

At this point, it is instructive to have a closer look at the
quadratic constraint. In three dimensions, this constraint implies
that the tensor
\begin{equation}
  \label{eq:Q1}
  {\cal Q}_{\mathcal{M}\llceil\mathcal{PQ}\rrfloor} ~\equiv~ 
  \Theta_{{\cal MN}}\,Z^{\cal N}{}_{\cal PQ} ~=~ -
  X_{\mathcal{M}\llceil\mathcal{P}}{}^\mathcal{N}\,
  \Theta_{\mathcal{Q}\rrfloor\mathcal{N}} \;,
\end{equation}
must vanish. Let us determine, in which representation
$\mathcal{Q}_{\mathcal{M}\llceil\mathcal{PQ}\rrfloor}$ transforms.  As
we have seen above, the tensor $Z^{\cal N}{}_{\cal PQ}$ in its indices
${\cal PQ}$ projects onto the ${\bf 3875}$ representation.  As a
consequence, ${\cal Q}_{\mathcal{M}\llceil\mathcal{PQ}\rrfloor}$
transforms in the tensor product ${\bf 248}\times{\bf 3875}$ given in
(\ref{eq:repQ1}).  On the other hand, as
$\mathcal{Q}_{\mathcal{M}\llceil\mathcal{PQ}\rrfloor}$ is quadratic in
$\Theta$ it transforms in the symmetric tensor product ${\bf 3875}
\times_{\rm sym} {\bf 3875}$ given in (\ref{eq:repQ2}).  Comparing
(\ref{eq:repQ2}) and (\ref{eq:repQ1}), it follows that also
$\mathcal{Q}_{\mathcal{M}\llceil\mathcal{PQ}\rrfloor}$ transforms in
the representation,
\begin{eqnarray}
C_{\rm quad} &=& {\bf 3875}+ {\bf 147250} \;,
\label{QQrep}
\end{eqnarray}
and thus in the very same representation as the
three-forms~(\ref{eq:Qrep}).  This is in accord with the general
pattern in gauged supergravities noted in the previous section: the
quadratic constraint transform in a (reducible) representation whose
conjugate is equal to (or at least contained in) the representation of
the $D$-forms.  We will propose a natural interpretation for this in
the last section, where the $D$-forms act as Lagrange multipliers for
the quadratic constraint.

Let us finally continue the tensor hierarchy one last step further,
{\it i.e.},\ to the four-forms. Although four-forms cannot live in
three dimensions, their tensor gauge freedom shows up in the
three-dimensional tensor gauge algebra by the shift transformation of
the three-forms (\ref{eq:delta-C-4}). For a complete picture we thus
need to work out also their structure.  Again, $Y^{\cal K\llceil
  MN\rrfloor}{}_{\cal P\llceil Q\llceil RS\rrfloor\rrfloor}$ does not
map onto the full tensor product ${\cal P\llceil Q\llceil
  RS\rrfloor\rrfloor}$ but only onto a restricted subrepresentation of
${\bf 248}\times ({\bf 3875}+{\bf 147250})$, which we do not
explicitly work out here.  It is interesting to note, that apart from
the standard othogonality relations~(\ref{eq:orthoY}) which follow as
a consequence of the quadratic constraint~(\ref{eq:XX-commutator}),
the tensor $Y^{\cal K\llceil MN\rrfloor}{}_{\cal P\llceil Q\llceil
  RS\rrfloor\rrfloor}$ also identically satisfies the relation
\begin{eqnarray}
  \label{eq:QY}
  {\cal Q}_{\mathcal{K}\llceil\mathcal{MN}\rrfloor} \,Y^{\cal K\llceil
    MN\rrfloor}{}_{\cal P\llceil 
    Q\llceil RS\rrfloor\rrfloor} &=&0\;,
\end{eqnarray}
with ${\cal Q}_{\mathcal{K}\llceil\mathcal{MN}\rrfloor}$ from
(\ref{eq:Q1}). This identity will also play an important role in the
last section. Its proof is not entirely straightforward, as (\ref{eq:QY}) 
involves expressions cubic in $\Theta$ and quadratic in the
$\mathrm{E}_8$ structure constants, and is therefore most easily
checked on a computer.

To summarize, we have explicitly worked out the tensor hierarchy of
gauged three-dimensional supergravity and shown that consistency
requires two- and three-forms to transform in the ${\bf 3875}$ and
${\bf 3875}+{\bf 147250}$ representation, respectively.  The
representation content of the (evanescent) four-forms is implicitly
defined by (\ref{eq:higherY-p=3}) as a subrepresentation of ${\bf
  248}\times ({\bf 3875}+{\bf 147250})$ and shows up through the shift
transformations~(\ref{eq:delta-C-4}) of the three-forms. In
principle, the precise representation content of the index
combinations in (\ref{eq:higherY-p=3}) can be worked out further, but
these details are not necessary in what follows.

%%%%%%%%%%%%%%%%%%%%%%%%%%%%%%%%%%%%%%%%%%%%%%%%%%%%%%%%%%%%%%%%%%
%%%%%%%%%%%%%%%%%%%%%%%%%%%%%%%%%%%%%%%%%%%%%%%%%%%%%%%%%%%%%%%%%%
\section{The supersymmetry algebra in three space-time dimensions}
\setcounter{equation}{0}
\label{sec:susy-algebra}
%%%%%%%%%%%%%%%%%%%%%%%%%%%%%%%%%%%%%%%%%%%%%%%%%%%%%%%%%%%%%%%%%%
In this section we present the complete determination of the
supersymmetry transformations and the corresponding algebra for the
$p$-forms in three dimensions.  Already in a number of cases
supersymmetry variations of $p$-forms that do not appear in the
ungauged action, have been determined. This was done by making an
ansatz for these variations based on their tensorial structure, which
involves some undetermined coefficients. These constants are
subsequently fixed by imposing the supersymmetry algebra, after which
one proceeds by iteration. Here we go one step further and consider
also the supersymmetry variations of those $p$-forms that are not
required for writing down the most general gaugings, in order to
determine what their possible role could be. In three space-time
dimensions this implies that we will now also consider the two-,
three-, and four-form potentials. Although four-form potentials do not
exist in four dimensions, their symmetries will still play a role as
they act on the three-form potentials. We note that a somewhat similar
investigation of maximal supergravity in five dimensions, in the
context of a framework based on $\mathrm{E}_{11}$, has recently
appeared in \cite{Riccioni:2007ni}.

We use spinor and ${\rm E}_{8(8)}$ conventions
from~\cite{Nicolai:2000sc,Nicolai:2001sv}.\footnote{%%%%%%%%%%%%%
  To be precise: the only change in notation with respect
  to~\cite{Nicolai:2000sc,Nicolai:2001sv} is the sign of the vector
  fields, {\it i.e.}, $A_{\mu}{}^\mathcal{M}\rightarrow
  -A_{\mu}{}^\mathcal{M}$. The tangent space metric and gamma matrix
  conventions are as follows: $\eta_{ab}= \mathrm{diag} (+,-,-)$,
  $\{\gamma^a,\gamma^b\} = 2\eta^{ab}\,\mathbf{1}$, and
  $\gamma^{abc}= - \mathrm{i}\varepsilon^{abc}\,\mathbf{1}$. } % 
%%%%%%%%%%%%%%%%%%%%%%%%%%%%%%%%%%%%%%%%%%%%%%%%  
In particular, the ${\rm E}_{8(8)}$ generators $t^{\cal M}$ split into
120 compact ones $X^{IJ}=X^{[IJ]}$, associated with the group
$\mathrm{SO}(16)$, and 128 non-compact ones denoted by $Y^{A}$. Here
$I,J,\dots$ and $A, B,\dots$, respectively, label the ${\bf 16}_v$ and
${\bf 128}_s$ representations of SO(16). Eventually we will also need
indices $\dot A,\dot B,\dots$ labelling the conjugate spinor
representation ${\bf 128}_c$. Naturally we will also encounter
$\mathrm{SO}(16)$ gamma matrices $\Gamma^I{}_{A\dot A}$ in what
follows. We will freely raise and lower $\mathrm{SO}(16)$ indices.

The scalar fields parametrize the ${\rm E}_{8(8)}/{\rm SO}(16)$ coset
space in terms of an ${\rm E}_{8(8)}$-valued matrix ${\cal V}^{\cal
  M}{}_{\underline{\cal P}}$, which transforms as
\begin{eqnarray}
  \label{eq:delta-V}
  \delta {\cal V}(x)^{\cal M}{}_{\underline{\cal P}} &=&
  - g_{\mathcal{N}}{}^{{\cal M}}\,{\cal V}(x)^{\cal
  N}{}_{\underline{\cal P}}+ 
  {\cal V}(x)^{\cal M}{}_{\underline{\cal Q}}
  \;h(x)_{\underline{\mathcal{P}}}{}^{\underline{\cal Q}} \;,  
\end{eqnarray}
under global ${\rm E}_{8(8)}$ and local ${\rm SO}(16)$, characterized
by the matrices $g$ and $h(x)$ which take their values in the Lie
algebra of the two groups. Note that underlined $\mathrm{E}_{8(8)}$
indices and indices $[IJ]$, $A$ and $\dot A$ are always subject to
{\it local} $\mathrm{SO}(16)$. The one-forms associated with the
scalars are given by
\begin{eqnarray}
  \label{eq:O+P}
  {\cal V}^{-1}\,D_{\mu}{\cal V}&=& \ft12\,{\cal
  Q}_{\mu}{}^{IJ}\,X^{IJ}+{\cal P}_{\mu}{}^{A}\,Y^{A}\;, 
\end{eqnarray}
where the derivative $D_\mu$ on the left-hand side is covariant with
respect to the chosen gauge group ({\it cf.}
(\ref{eq:vector-gauge-tr})). As is well-known, $\mathcal{Q}_\mu{}^{IJ}$
will play the role of a composite $\mathrm{SO}(16)$ gauge connection.
Both $\mathcal{P}_\mu$ and $\mathcal{Q}_\mu$ will implicitly depend on
the gaugings introduced in section \ref{sec:hierarchy}, through the
defining relation (\ref{eq:O+P}).

For simplicity of the formulas we use the abbreviating notation,
\begin{eqnarray}
  \label{eq:abbreviation-V}
  {\cal V}^{\cal MN}{}_{\underline{\cal P}|\underline{\cal R}} &\equiv&
  {\cal V}^{\cal M}_{\;\;\;\;\underline{\cal P}}\,{\cal V}^{\cal
    N}_{\;\;\;\underline{\cal R}} \;, \quad
  {\cal V}^{\cal MNK}{}_{\underline{\cal P}|\underline{\cal
      R}|\underline{\cal S}} ~\equiv~ 
  {\cal V}^{\cal M}_{\;\;\;\;\underline{\cal P}}\,{\cal V}^{\cal
    N}_{\;\;\;\underline{\cal R}} 
  \,{\cal V}^{\cal K}_{\;\;\;\underline{\cal S}} \;,
  \quad {\rm etc.}
\end{eqnarray}
for multiple tensor products of these matrices.  The fermionic field
content is given by 16 gravitinos $\psi_\mu{}^I$ and 128 spin-1/2
fermions $\chi^{\dot A}$ transforming under ${\rm SO}(16)$. In the
presence of a gauging their supersymmetry variations are given by
\begin{eqnarray}
  \label{eq:susyfermions}
  \delta\,\psi_\mu{}^I &=&  D_{\mu}\epsilon^I 
  + \mathrm{i} g\,{A}_{1}{}^{IJ}\,\gamma_{\mu} \epsilon^J 
  \;,\qquad
  \delta\,\chi^{\dot A} ~=~
  \ft12{\mathrm{i}}\,\gamma^\mu\epsilon^I\,\Gamma^I_{A\dot A}\,{\cal P}_\mu^A 
  +g\,A_{2}{}^{I\dot A}\,\epsilon^I \;,
\end{eqnarray}
with the tensors $A_{1}$, $A_{2}$ given by
\begin{eqnarray}
  \label{eq:def-A1-A2} 
  A_{1}^{IJ} &=& \ft1{7}\,
  {\cal V}^{\cal MN}{}_{IK|JK}\,
  \Theta_{{\cal MN}}\;,
  \qquad
  A_{2}^{I\dot A}~=~
  -\ft17\,\Gamma^J_{A\dot A}\,
  {\cal V}^{\cal MN}{}_{IJ|A}\,
  \Theta_{{\cal MN}}  \;.
\end{eqnarray}
The bosonic fields on the other hand transform as
\begin{eqnarray}
  \label{eq:susybosons}
  \delta {e_\mu}^\alpha &=& 
  \mathrm{i} \bar{\epsilon}^I \gamma^\alpha \psi_\mu{}^I 
  \;,\qquad
  {\cal V}^{-1} \delta {\cal V} ~=~ 
  \Gamma^I_{A\dot A} \,\bar{\chi}^{\dot A} \epsilon^I \,Y^A
  \;,\nonumber\\[.5ex]
  \delta A_\mu{}^{\cal M} &=&
  2 \,{\cal V}^{\cal M}_{\;\;\;\;IJ}\,{\bar{\epsilon}}^I\psi_\mu{}^J 
  -\mathrm{i} \Gamma^I{}_{A\dot A}\,{\cal V}^{\cal
    M}_{\;\;\;\;A}\;{\bar{\epsilon}}^I\gamma_{\mu}\chi^{\dot A}  
  \;.
\end{eqnarray}

The supersymmetry transformations are expected to close into the
various local symmetries, up to field equations. The supersymmetry
comutator takes the form,
\begin{eqnarray}
  \label{eq:susyalgebra}
  {}[\delta_{\epsilon_{1}},\delta_{\epsilon_{2}}] &=&
  \xi^\mu\hat D_\mu+\delta_{\Lambda}+\delta_{\Xi}+\delta_{\Phi} +
  \cdots \;, 
\end{eqnarray}
where the unspecified terms denote local Lorentz transformations,
local supersymmetry transformations and other symmetries which will be
discussed below. By $\xi^\mu\hat D_\mu$ we denote a covariant
translation: a general coordinate transformation with parameter
$\xi^\mu$ accompanied by other field-dependent gauge transformations
such that the combined result is fully covariant. In the context of
this work we are mostly interested in the field-dependent vector and
tensor gauge transformations, 
\begin{eqnarray}
  \label{eq:covariant-translation}
  \xi^\mu \hat D_\mu &\equiv& \xi^\mu\partial_\mu +\delta_{\Lambda(\xi)}+
\delta_{\Xi(\xi)}+\delta_{\Phi(\xi)} + \cdots  \;,
\end{eqnarray}
where the vector and tensor gauge parameters are equal to
\begin{eqnarray}
  \label{eq:cov-translation-parameters} 
  \Lambda(\xi)^{\cal M} &\equiv& -\xi^{\rho}A_{\rho}{}^{\cal M}\;,
  \nonumber\\ 
  \Xi(\xi)_{\mu}{}^{\cal MN} &\equiv& -\xi^{\rho}\,\Big(
  B_{\rho\mu}{}^{\cal MN}
  + A_{\rho}{}^{\cal \llceil M} A_{\mu}{}^{\cal N\rrfloor}\Big)\;,
  \nonumber\\ 
  \Phi(\xi)_{\mu\nu}{}^{\mathcal{K}\llceil\mathcal{MN}\rrfloor} &\equiv&
  -\xi^{\rho}\,\Big( 
  C_{\rho\mu\nu}{}^{\mathcal{K}\llceil\mathcal{MN}\rrfloor} 
  - A_{\rho}{}^{\cal \llceil K} B_{\mu\nu}{}^{\cal MN\rrfloor}
  -\ft23\, A_{[\mu}{}^{\cal \llceil K}  A_{\nu]}{}^{\cal \llceil M} 
  A^{\vphantom{\llceil}}_{\rho}{}^{\cal N\rrfloor\rrfloor}  
  \Big) \;,
\end{eqnarray}
so that 
\begin{eqnarray}
  \xi^\rho \hat D_\rho A_{\mu}{}^{\cal M} &=& 
  \xi^{\rho}\,{\cal H}_{\rho\mu}{}^{\cal M}\;,\nonumber\\
  \xi^\rho\hat D_\rho B_{\mu\nu}{}^{\cal M\cal N} -
   A_{\mu}{}^{\cal \llceil M} \xi^\rho\hat D_{\rho} 
  A_{\nu}{}^{\cal N\rrfloor} + A_{\nu}{}^{\cal \llceil M}
  \xi^\rho\hat D_{\rho}  
  A_{\mu}{}^{\cal N\rrfloor}  
  &=& \xi^{\rho}\,{\cal H}_{\rho\mu\nu}{}^{\cal MN}\;,
\end{eqnarray}
take a fully covariant form in terms of the covariant variations and
field strengths of section~\ref{sec:hierarchy}. Note that we have
suppressed the supercovariantizations in this result, as we restrict
attention to the terms of lowest order in the fermion fields.
Calculating closure of the supersymmetry algebra on the $p$-form
tensor fields will determine the parameters $\xi$, $\Lambda$, $\Xi$,
$\Phi$ in (\ref{eq:susyalgebra}).

Let us start from the supersymmetry commutator on vector fields. A
short computation starting from (\ref{eq:susyfermions}) and
(\ref{eq:susybosons}) yields
\begin{eqnarray}
  \label{eq:commA1}
  {}[\delta_{\epsilon_{1}},\delta_{\epsilon_{2}}]\,A_{\mu}{}^{\cal M}
  &=& 
  -2D_{\mu}({\cal V}^{\cal
    M}_{\;\;\;\;IJ}\,\bar\epsilon_{[1}{}^I\epsilon_{2]}{}^{J}) 
  +\mathrm{i}\,\epsilon_{\mu\nu\rho}\,{\cal V}^{\cal M}_{\;\;\;\;A}\,
  {\cal P}^{\nu\,A}\,
  \bar\epsilon_{[1}{}^{I}\gamma^{\rho}\epsilon_{2]}{}^{I}
  \nonumber\\
  &&{}
  +2\mathrm{i} g\,\Big(
  \Gamma^{I}_{A\dot A}\,A_{2}{}^{J\dot A}\,{\cal V}^{\cal M}_{\;\;\;\;A}-
  2A_{1}{}^{JK}\,{\cal V}^{\cal M}_{\;\;\;\;IK}\Big)\,
  \bar\epsilon_{[1}{}^{I}\gamma_{\mu}\epsilon_{2]}{}^{J}   \;.
\end{eqnarray}
The first term is a gauge transformation, while the last term proves
to be the dressed version of the constant tensor $Z^{\cal M}{}_{\cal
  PQ}$ defined in (\ref{eq:def-Z}). Indeed, we note the identity,
\begin{eqnarray}
  \label{eq:gamma-A2-V}
  \Gamma^{(I}_{A\dot A}\,A_{2}{}^{J)\dot A}\,{\cal V}^{\cal M}_{\;\;\;\;A}-
  A_{1}{}^{JK}\,{\cal V}^{\cal M}_{\;\;\;\;IK} -
  A_{1}{}^{IK}\,{\cal V}^{\cal M}_{\;\;\;\;JK}
  &=&\ft27\,{\cal V}^{\cal PQ}{}_{IK|JK}\,
  Z^{\cal M}{}_{\cal PQ}\;.
\end{eqnarray}
Upon contraction with $\Theta_{\cal{MN}}$, the right-hand side of
this equation vanishes, and we re-obtain the identity (3.18) of
\cite{Nicolai:2001sv}.  The second term in (\ref{eq:commA1}) shows up in
the duality equation relating vector and scalar fields in three
dimensions,
\begin{eqnarray}
  \label{eq:dualvectors}
  {\cal X}_{\mu\nu}{}^{\cal M} &\equiv& 
  {\cal H}_{\mu\nu}{}^{\cal M}
  + e \,\epsilon_{\mu\nu\rho}\,{\cal V}^{\cal M}_{\;\;\;\;A}\, {\cal
    P}^{\rho\, A} \;,
\end{eqnarray}
which, at least in the ungauged theory, vanishes on-shell. Hence, we
find that
\begin{eqnarray}
  \label{eq:ddA}
  {}[\delta_{\epsilon_{1}},\delta_{\epsilon_{2}}]\,A_{\mu}{}^{\cal M} &=&
  \xi^{\rho}\,{\cal H}_{\rho\mu}{}^{\cal M}+
  D_{\mu}\,\Lambda^{\cal M}
  -g\,Z^{\cal M}{}_{\cal PQ}\,\Xi_{\mu}{}^{\cal PQ}
  -\xi^{\rho}\,{\cal X}_{\rho\mu}{}^{\cal M} \;,
\end{eqnarray}
with parameters
\begin{eqnarray}
  \label{susypars}
  \xi^{\mu}&=&-\mathrm{i}\,
  \bar\epsilon_{[1}{}^I\gamma^{\mu}\epsilon_{2]}{}^{I} \;,
  \nonumber\\[1ex]
  \Lambda^{\cal M}&=&
  -2\,{\cal V}^{\cal M}_{\;\;\;\;IJ}\,
  \bar\epsilon_{[1}{}^{I}\epsilon_{2]}{}^J \;,
  \nonumber\\[1ex]
  \Xi_{\mu}{}^{\cal MN}&=&
  -\ft47 \mathrm{i}\,{\cal V}^{\cal \llceil MN\rrfloor}{}_{IK|JK}\:
  \bar\epsilon_{[1}{}^{I}\gamma_{\mu}\epsilon_{2]}{}^{J} \;.
\end{eqnarray}
Except for the last term in~(\ref{eq:ddA}) the supersymmetry algebra
closes precisely as expected. Usually, this last term is disregarded
as the supersymmetry algebra is expected to close modulo the
first-order (duality) equations of motion (that is, ${\cal
  X}_{\mu\nu}{}^{\cal M} =0$). Nowever, matters are more subtle here,
as only a projection of the duality equation with the embedding tensor
is expected to correspond to an equation of motion. For the moment,
let us just keep this term: we will interpret it later as an
additional local symmetry of the Lagrangian.

Let us continue with the two-forms. The supersymmetry variation of
$B_{\mu\nu}{}^{\cal MN}$ is determined by its tensor structure up to
two constants, $\alpha_{1}$ and $\alpha_{2}$, 
\begin{eqnarray}
  \label{eq:susyB}
  \Delta B_{\mu\nu}{}^{\cal MN} &=& \mathrm{i}\alpha_{1} \, {\cal V}^{\cal
  \llceil MN\rrfloor}{}_{IK|JK}\;
  {\bar{\epsilon}}^I\gamma_{[\mu}\psi_{\nu]}{}^J 
-\alpha_{2}\, {\cal V}^{\cal \llceil MN\rrfloor}{}_{A|IJ}\;
\Gamma^{I}_{A\dot A}\,{\bar{\epsilon}}^J\gamma_{\mu\nu}\chi^{\dot A}
\;.
\end{eqnarray}
Requiring that the commutator closes into a gauge transformation with
parameter $\Xi_{\mu}{}^{\cal MN}$ as given in (\ref{susypars}), leads
to $\alpha_{1}=-8/7$, $\alpha_{2}=-4/7$. From (\ref{eq:susyB}), we
obtain after some further computation,
\begin{eqnarray}
  \label{ddB1}
  {}[\delta_{\epsilon_{1}},\delta_{\epsilon_{2}}]\,B_{\mu\nu}{}^{\cal
    MN} 
  &=&{}
  2D_{[\mu}\Xi_{\nu]}{}^{\cal MN}
  +\ft47\,\epsilon_{\mu\nu\rho}\,{\cal P}^{\rho\,B}\,
  {\cal V}^{\cal \llceil MN\rrfloor}{}_{IJ|A}\,
  (\Gamma^{I}\Gamma^{K})_{AB}\;\bar\epsilon_{[1}{}^{J}\epsilon_{2]}{}^{K} 
  \nonumber\\[.5ex]
  &&{}
  -\ft87\,g\,
  \Big(
  {\cal V}^{\cal \llceil MN\rrfloor}{}_{IK|LK}\,A_{1}{}^{JL}
  +\ft12
  {\cal V}^{\cal \llceil MN\rrfloor}{}_{IK|A}\,\Gamma^{K}{}_{A\dot
    A}\,A_{2}{}^{J\dot A} 
  \Big)\;
  \bar\epsilon_{[1}{}^{I}\gamma_{\mu\nu}\epsilon_{2]}{}^{J} 
  \nonumber\\[.5ex]
  &&{}
  +2\, A_{[\mu}{}^{\cal \llceil
    M}\,[\delta_{\epsilon_{1}},\delta_{\epsilon_{2}}]\, A_{\nu]}{}^{\cal
    N\rrfloor} \;.
\end{eqnarray}
The first term denotes the tensor gauge transformation. To understand
the second term we need to make explicit use of the projection of
${\cal \llceil MN\rrfloor}$ onto the ${\bf 3875}$, which induces
relations such as ~\cite{Nicolai:2000sc,Nicolai:2001sv},
\begin{eqnarray}
  \label{eq:V-projection} 
  {\cal V}^{\cal \llceil MN\rrfloor}{}_{IJ|A}&=&
  \ft1{14}\,\Big(
  (\Gamma^{I}\Gamma^{K})_{AB}\,{\cal V}^{\cal \llceil MN\rrfloor}{}_{KJ|B}-
  (\Gamma^{J}\Gamma^{K})_{AB}\,{\cal V}^{\cal \llceil MN\rrfloor}{}_{KI|B}
  \Big)\;.
\end{eqnarray}
After some calculation, the second term in (\ref{ddB1}) then reduces
to $2\,\Lambda^{\cal \llceil M}\,( {\cal X}_{\mu\nu}{}^{\cal
  N\rrfloor} - {\cal H}_{\mu\nu}{}^{\cal N\rrfloor})$, where we again
introduced the expression for the duality
relation~(\ref{eq:dualvectors}). The term proportional to
$\mathcal{H}_{\mu\nu}{}^\mathcal{N}$ then yields a term belonging to
the tensor gauge transformation (\ref{eq:Delta-gauge-B2}). The second
line in (\ref{ddB1}) can be simplified in a similar way.  Its $(IJ)$
traceless part may be brought into the form
\begin{eqnarray} 
  \label{eq:IJ-traceless}
  \ft1{7}\,g\,Y^{\cal
    MN}{}_{{\cal K\llceil PQ\rrfloor}}\, {\cal V}^{\llceil\mathcal{K}\llceil
    \mathcal{PQ}\rrfloor\rrfloor}{}_{IK|KL|LJ}\;
  \bar\epsilon_{[1}{}^{I}\gamma_{\mu\nu}\epsilon_{2]}{}^{J} \;,
\end{eqnarray}
and thus constitutes the shift transformation of
(\ref{eq:Delta-gauge-B2}) with parameter\footnote{%%%%%%%%%%%%%%%
  Note that not only the coefficient is determined. There exists yet
  another independent term with the correct tensor structure,
  $\Gamma^{IK}_{AB}\,{\cal V}^{\cal \llceil K\llceil
    MN\rrfloor\rrfloor}{}_{A|B|JK}\;
  \bar\epsilon_{[1}{}^{I}\gamma_{\mu\nu}\epsilon_{2]}{}^{J}$, which
  turns out to be absent. One may verify by explicit calculation that
  $\Phi^{\cal K\llceil MN\rrfloor}$ defined in (\ref{eq:defPhi}) has
  contributions in both irreducible representations ${\bf 3875}$ and
  ${\bf 147250}$.  }
%%%%%%%%%%%%%%%%%%%%%%%%%%%%%%%%%%%%%%%%%%%%%%%%%%%%%%%%%
\begin{eqnarray}
  \label{eq:defPhi}
  \Phi_{\mu\nu}{}^{\cal K\llceil MN\rrfloor}&=& -\ft17\,
  {\cal V}^{\cal \llceil K\llceil MN\rrfloor\rrfloor}{}_{IK|KL|LJ}\;
  \bar\epsilon_{[1}{}^{I}\gamma_{\mu\nu}\epsilon_{2]}{}^{J}\;.
\end{eqnarray}
It remains to consider the $(IJ)$ trace part of the second line in
(\ref{ddB1}) which reduces to
\begin{eqnarray}
  \label{eq:epsilon-V-eq}
  4\,e g\,\varepsilon_{\mu\nu\rho}\,\xi^{\rho}\;{\mathbb V}^{\cal \llceil
    MN\rrfloor,\llceil KL\rrfloor}\, \Theta_{\cal KL}
  \;,
\end{eqnarray}
where ${\mathbb V}^{\cal \llceil MN\rrfloor,\llceil KL\rrfloor}$ equals the
symmetric scalar-dependent matrix defined by
\begin{eqnarray} 
  \label{eq:defVV}
  {\mathbb V}^{\cal \llceil MN\rrfloor,\llceil KL\rrfloor}&=&
  \ft1{392}\,\Big(
  7\,{\cal V}^{\cal \llceil MN\rrfloor\llceil KL\rrfloor}{}_{IJ|A|IJ|A}
  -2\,{\cal V}^{\cal \llceil MN\rrfloor\llceil KL\rrfloor}{}_{IK|JK|IL|JL}
  \Big)  \;.
\end{eqnarray}
Putting everything together, the supersymmetry commutator on two-forms
takes the form, 
\begin{eqnarray}
  \label{eq:ddB}
  {}[\delta_{\epsilon_{1}},\delta_{\epsilon_{2}}]\,B_{\mu\nu}{}^{\cal
    MN} &=& 
  4ge\,\xi^{\rho}\,\varepsilon_{\mu\nu\rho}\,{\mathbb V}^{\cal\llceil
    MN\rrfloor,\llceil KL\rrfloor}\,\Theta_{\cal KL} 
  +2D_{[\mu}\Xi_{\nu]}{}^{\cal MN}
  -2\,{\cal H}_{\mu\nu}{}^{\cal \llceil M}\,\Lambda^{\cal N\rrfloor}
  \nonumber\\[.4ex]
  &&{}
  -gY^{\cal MN}{}_{\cal K\llceil PQ\rrfloor}\,\Phi_{\mu\nu}{}^{\cal
    K\llceil PQ\rrfloor} 
  +2\,{\cal X}_{\mu\nu}{}^{\cal \llceil M}\,\Lambda^{\cal N\rrfloor}
  +2\,A_{[\mu}{}^{\cal \llceil
    M}\,[\delta_{\epsilon_{1}},\delta_{\epsilon_{2}}]\, A_{\nu]}{}^{\cal
    N\rrfloor} \;.
  \nonumber\\
  &=&
  \Big(\xi^\rho \hat D_\rho
  +\delta_{\Lambda}+\delta_{\Xi}+\delta_{\Phi}\Big)\, 
  B_{\mu\nu}{}^{\cal MN} -\xi^{\rho}{\cal Y}_{\rho\mu\nu}{}^{\cal
    MN}
  \nonumber\\[.5ex]
  &&{} +2\,{\cal X}_{\mu\nu}{}^{\cal \llceil M}\,(\Lambda^{\cal
    N\rrfloor}+\Lambda(\xi)^{\cal N\rrfloor}) -2\,\xi^{\rho}\,{\cal
    X}_{\rho[\mu}{}^{\cal \llceil M}A_{\nu]}{}^{\cal N\rrfloor} \;, 
\end{eqnarray}
where in the second equation we introduced the tensor, 
\begin{eqnarray}
  \label{eq:duals}
  {\cal Y}_{\mu\nu\rho}{}^{\cal MN}
  &\equiv&
  {\cal H}_{\mu\nu\rho}{}^{\cal MN}
  -4\,g\,e\,\varepsilon_{\mu\nu\rho}\,{\mathbb V}^{\cal\llceil
    MN\rrfloor,\llceil KL\rrfloor}\,\Theta_{\cal KL} 
  -6\, 
  A_{[\mu}{}^{\cal \llceil M}\,{\cal X}_{\nu\rho]}{}^{\cal N\rrfloor}
  \;. 
\end{eqnarray} 
This tensor takes the form of a duality relation between the field
strength of the two-forms (\ref{eq:H-3-cov}) and the embedding tensor.
The supersymmetry commutator thus closes according
to~(\ref{eq:susyalgebra}) modulo terms proportional to the duality
relations~(\ref{eq:dualvectors}) and (\ref{eq:duals}). These terms are
interpreted as follows. The term proportional to
$\mathcal{Y}_{\mu\nu\rho}{}^\mathcal{MN}$ corresponds to a new
symmetry transformation of the two-form potential. The last term
proportional to $\mathcal{X}_{\rho\mu}{}^\mathcal{M}$ accompanies the
extra transformation in the vector fields represented by the last term
in (\ref{eq:ddA}). Finally the preceding terms proportional to
$\mathcal{X}_{\mu\nu}{}^\mathcal{M}$ are interpreted as deformations
of the vector gauge transformation acting on the two-form
potential ({\it cf.} (\ref{eq:gauge-tr-ABC})). Hence we change this
transformation according to, 
\begin{eqnarray} 
  \label{eq:deformed-Lambda-B}
    \delta_\mathrm{mod}(\Lambda) B_{\mu\nu}{}^{\mathcal{MN}}  
  &=&{} -2\,
  \Lambda^{\llceil\mathcal{M}}\mathcal{H}_{\mu\nu}{}^{\mathcal{N}\rrfloor} 
  +2\,
  \Lambda^{\llceil\mathcal{M}}\mathcal{X}_{\mu\nu}{}^{\mathcal{N}\rrfloor}  
  + 2\, A_{[\mu}{}^{\llceil\mathcal{M}}\,
  \delta(\Lambda)A_{\nu]}{}^{\mathcal{N}\rrfloor}  \nonumber\\
  &=&{}
  2 e\,\varepsilon_{\mu\nu\rho}\,
  \Lambda^{\llceil\mathcal{M}}\mathcal{V}^{\mathcal{N}\rrfloor}{}_A\;
  \mathcal{P}^{\rho A} 
  + 2\, A_{[\mu}{}^{\llceil\mathcal{M}}\,
  \delta(\Lambda)A_{\nu]}{}^{\mathcal{N}\rrfloor} \,. 
\end{eqnarray}
This deformation is reminiscent of what happens, for instance, in
$D=4$ gauged supergravity
\cite{deWit:2005ub,deVroome:2007zd,deWit:2007mt}, where the two-form
fields acquire also additional variations once they couple to other
fields in the Lagrangian. Of course, it remains to see whether this
interpretation is correct, but we will present further evidence of
this in section \ref{sec:Lagrangian-forms-3d}.

The duality relation~(\ref{eq:duals}) is remarkable. On-shell, 
({\it i.e.}\ for ${\cal X}^{\cal M}=0={\cal Y}^{\cal MN}$) it reads
\begin{eqnarray}
  \label{eq:duals1}
{\cal H}_{\mu\nu\rho}{}^{\cal MN} =4g\,e\,\varepsilon_{\mu\nu\rho}\,
{\mathbb V}^{\cal\llceil MN\rrfloor,\llceil KL\rrfloor}\,\Theta_{\cal KL}
\;,
\end{eqnarray}
and it relates the field strengths of the two-forms to the embedding
tensor.  The scalar matrix ${\mathbb V}^{\cal\llceil
  MN\rrfloor,\llceil KL\rrfloor}$ defined in (\ref{eq:defVV}), which
shows up in this equation, is related to the scalar potential of the
gauged theory in a simple way.  With the explicit expression for the
scalar potential $V$ from~\cite{Nicolai:2000sc,Nicolai:2001sv} one
finds the expression
\begin{eqnarray}
  \label{eq:Pot}
  V &=&
  -\ft18\Big( A_{1}^{IJ}A_{1}^{IJ}-\ft12\, 
  A_{2}^{I\dot A}A_{2}^{I\dot A} \Big)
  ~=~
  \ft12\,{\mathbb V}^{\cal\llceil MN\rrfloor,\llceil KL\rrfloor}\,
  \Theta_{\cal MN}\,\Theta_{{\cal KL}}
  \;.
\end{eqnarray}
In other words, the matrix ${\mathbb V}^{\cal\llceil
  MN\rrfloor,\llceil KL\rrfloor}$ precisely encodes the scalar
potential of the gauged theory.  This appears to be a generic pattern
for the $(D\!-\!1)$-forms in the gauged supergravities, and we shall
see its natural interpretation in the next section. We emphasize that
the matrix ${\mathbb V}^{\cal\llceil MN\rrfloor,\llceil KL\rrfloor}$
is {\em not} positive definite --- unlike the scalar matrices that
show up in the lower-rank $p$-form dualities.  This lack of positivity
is in accord with the fact that the potentials of gauged
supergravities are generically known to be unbounded from below.

At this point let us briefly comment on a similar result in
\cite{Riccioni:2007ni} where the form fields are considered for $D=5$
gauge maximal supergravity. In that work an equation (4.27) appears
which seems the direct analogue of (\ref{eq:duals1}), but now for the
field strength of the four-form potential. Although it has the same
structure as (\ref{eq:duals1}), its right-hand side is not related to
the potential in the way we described above. However, a direct
comparison is subtle as (\ref{eq:dualvectors}) only vanishes on shell
upon projection with the embedding tensor, so that (\ref{eq:duals1})
will not be realized as a field equation.

The duality equation~(\ref{eq:duals1}) in particular provides
the ${\rm E}_{8(8)}$ covariant field equation for two-forms in the
three-dimensional ungauged theory:
\begin{eqnarray}
  \label{field2}
  \partial^{\mu}\Big({\mathbb V}_{\cal\llceil MN\rrfloor,\llceil
    KL\rrfloor}\, {\cal H}_{\mu\nu\rho}{}^{\mathcal{KL}} 
  \Big)~+~{\rm fermions} &=& 0\;,
\end{eqnarray}
with ${\mathbb V}_{\cal\llceil MN\rrfloor,\llceil KL\rrfloor}$ the
inverse matrix to ${\mathbb V}^{\cal\llceil MN\rrfloor,\llceil
  KL\rrfloor}$.

To close this section, we also compute the commutator of
supersymmetry transformations on the three-forms.
Equation~(\ref{eq:defPhi}) suggests to define the supersymmetry variation
of the three-forms as
\begin{eqnarray}
  \Delta C_{\mu\nu\rho}{}^{\cal K\llceil MN\rrfloor}
  &=&
  \ft37\, {\cal V}^{\cal \llceil K\llceil
    MN\rrfloor\rrfloor}{}_{IK|KL|LJ}\; 
  {\bar{\epsilon}}^I\gamma^{\vphantom{I}}_{[\mu\nu}\psi_{\rho]}{}^J 
  ~+~\cdots   \;,
\end{eqnarray}
where the dots refer to the $\bar\epsilon\,\chi$ variations. Indeed,
\begin{eqnarray}
  2\delta_{\epsilon_{[1}}
  \Big(\ft37\,
  {\cal V}^{\cal \llceil K\llceil MN\rrfloor\rrfloor}{}_{IK|KL|LJ}\;
  {\bar{\epsilon}_{2]}}^I\gamma_{[\mu\nu}\psi^J_{\rho]}
  \Big)
  &=&
  3\,D_{[\mu}\Phi_{\nu\rho]}{}^{\cal KMN}
  \\
  &&{}
  +\ft3{7}\,D_{[\mu}
  \Big({\cal V}^{\cal \llceil K\llceil MN\rrfloor\rrfloor}{}_{IK|KL|LJ}\Big)\;
  \bar\epsilon_{[1}{}^{I}\gamma_{\mu\nu}\epsilon_{2]}{}^{J} 
  + \cdots\;,
  \nonumber
\end{eqnarray}
thus reproducing the correct $\Phi$ term given in (\ref{eq:defPhi}).
Evaluating the derivatives of the second term and using the duality
equation~(\ref{eq:dualvectors}), eventually brings this term into the
form (modulo ${\cal X}^{[\cal M]}$),
\begin{eqnarray}
  \label{ddC1}
  &&
  \Big(\ft3{14}\,
  {\cal V}^{\cal \llceil K\llceil MN\rrfloor\rrfloor}{}_{A|KM|JM}-
  \ft3{7}\,
  {\cal V}^{\cal \llceil K\llceil MN\rrfloor\rrfloor}{}_{JM|KM|A}
  \Big)\,
  (\Gamma^{K}\Gamma^{N})_{AB}\,
  {\cal P}^{B}_{[\mu}\,
  \bar\epsilon^{N}_{[1}\gamma_{\nu\rho]}\epsilon^{J}_{2]}
  \nonumber\\
  &&
  {}
  \hspace{2cm}
  +3\,{\cal H}_{[\mu\nu}{}^{\cal \llceil K} \,
  \Xi_{\rho]}^{\cal MN\rrfloor}   \;.
\end{eqnarray}
In order to arrive at this result, we need to make use of the explicit
projection onto the ${\bf 3875}+{\bf 147250}$ within the tensor
product ${\cal \llceil K\llceil MN\rrfloor\rrfloor}$.  This gives rise
to a number of non-trivial identities, like
\begin{eqnarray}
  \label{eq:projection-id}
   14\,\Gamma^{K}_{A\dot A}
   {\cal V}^{\cal \llceil K\llceil
   MN\rrfloor\rrfloor}_{\;\;\;\;\;\;\;\; A|KM|JM}+
   16\,\Gamma^{K}_{A\dot A}
   {\cal V}^{\cal \llceil K\llceil
   MN\rrfloor\rrfloor}_{\;\;\;\;\;\;\;\; JM|KM|A}&&\nonumber\\
      - (\Gamma^{K}\Gamma^{MN})_{A\dot A}
   {\cal V}^{\cal \llceil K\llceil
   MN\rrfloor\rrfloor}_{\;\;\;\;\;\;\;\; MN|JK|A}
   &=& 0\;,  
\end{eqnarray}
which results from the projection of a triple product of ${\cal V}$'s
onto the $\bf{147250} + \bf{3875}$ representation in the same way as
(\ref{eq:V-projection}) is obtained by applying (\ref{eq:proj-3875})
to a double product of ${\cal V}$'s. {}From~(\ref{ddC1}) we can infer
the full supersymmetry transformation of the three-forms.  While the
last term is precisely expected from the tensor gauge transformations
(\ref{eq:Delta-gauge-B2}), the rest must be cancelled by
$\delta\chi$ variations in $\delta C$.  Together, this determines the
supersymmetry variation of the three-forms to be given by 
\begin{eqnarray}
  \Delta C_{\mu\nu\rho}{}^{\cal K\llceil MN\rrfloor}
  &=&
  \ft37\,
  {\cal V}^{\cal \llceil K\llceil MN\rrfloor\rrfloor}{}_{IK|KL|LJ}\;
  {\bar{\epsilon}}^I\gamma_{[\mu\nu}\psi^J_{\rho]}
  \label{susyC}
  \\
  &&{}-\ft1{14}\mathrm{i}\,
  \Big(
  {\cal V}^{\cal \llceil K\llceil MN\rrfloor\rrfloor}{}_{A|KM|JM}
  -
  2{\cal V}^{\cal \llceil K\llceil MN\rrfloor\rrfloor}{}_{JM|KM|A}
  \Big)\,
  \Gamma^{K}{}_{A\dot A}\, 
  {\bar{\epsilon}}^J\gamma_{\mu\nu\rho}\chi^{\dot A} 
  \;.
  \nonumber
\end{eqnarray}

To summarize, we have determined the supersymmetry variations of all $p$-forms
in three dimensions by closure of the supersymmetry algebra.
The full algebra is given by
\begin{eqnarray}
  \label{eq:susyalgebra1}
  {}[\delta_{\epsilon_{1}},\delta_{\epsilon_{2}}] &=&
  \xi^\mu\hat D_\mu+\delta_{\Lambda}+\delta_{\Xi}+\delta_{\Phi} + 
  \delta_{\cal X}+\delta_{\cal Y}  \;,
\end{eqnarray}
up to supersymmetry and local Lorentz symmetry transformations.  The
last two terms correspond to additional local symmetries proportional
to $\mathcal{X}_{\mu\nu}$ and $\mathcal{Y}_{\mu\nu\rho}$, that have
appeared in (\ref{eq:ddA}) and (\ref{eq:ddB}) for the one- and
two-forms, respectively. Furthermore, we recall that we have made a
modification in the vector gauge transformation rule for the
two-forms.

Of course, we have to justify both the presence of this deformation
and the fact that the two new variations can indeed be regarded as
symmetries of a specific Lagrangian. In this respect it is important
to recall that $\mathcal{X}_{\mu\nu}$ and $\mathcal{Y}_{\mu\nu\rho}$
take the form of first-order duality equations between $p$-forms in
three dimensions and, as it turns out, there are indeed field
equations proportional to $\mathcal{X}_{\mu\nu}$ and
$\mathcal{Y}_{\mu\nu\rho}$. This feature plays an important role in
realizing the invariance.  To understand this issue further we turn to
the construction of the Lagrangian in the next section.

%%%%%%%%%%%%%%%%%%%%%%%%%%%%%%%%%%%%%%%%%%%%%%%%%%%%%%%%%%%%%%%
\section{The Lagrangian with all $p$-forms in three dimensions}
\label{sec:Lagrangian-forms-3d} 
\setcounter{equation}{0}
%%%%%%%%%%%%%%%%%%%%%%%%%%%%%%%%%%%%%%%%%%%%%%%%%%%%%%%%%%%%%%%
Finally, we give a Lagrangian which contains all $p$-forms in three
dimensions. To this end we start from the gauged Lagrangian
of~\cite{Nicolai:2000sc,Nicolai:2001sv}, 
\begin{eqnarray}
  \label{eq:Lgauged}  
  {\cal L}_{\rm gauged} &=& 
  -\ft14 e R
  + \ft14 e {\cal P}^{\mu A} {\cal P}^A_\mu
  +\ft12\, \varepsilon^{\mu\nu\rho} \bar{\psi}{}^I_\mu D_\nu \psi_\rho^I 
  -\ft{1}{2}\mathrm{i}e \bar{\chi}^{\dot A} 
  \gamma^\mu D_\mu \chi^{\dot A} 
  \nonumber\\[1ex]
  &&{}
  -\ft14\,g\,\varepsilon^{\mu\nu\rho}\,A_\mu{}^{\cal M}\,\Theta_{\cal MN}\,
  (\partial_\nu A_\rho{}^{\cal N}
  +\ft13\,gX_{\cal [RS]}{}^{\cal N}\,A_\nu{}^{\cal R} A_\rho{}^{\cal S} ) 
  \nonumber\\[1ex]
  &&{}
  -\ft12e\,  \bar{\chi}^{\dot A} 
  \gamma^\mu \gamma^\nu \psi^I_\mu \,\Gamma^I{}_{A\dot A} {\cal P}^A_\nu 
  +\ft12eg\,
  A_{1}{}^{IJ}\;{\bar\psi}{}^I_{\mu}\,\gamma^{\mu\nu}\,\psi^{J}_{\nu} + 
  {\mathrm{i}}eg\,A_{2}{}^{I\dot A}\;
  {\bar\chi}{}^{\dot A}\gamma^\mu\,\psi^I_{\mu}
  \nonumber\\[1ex]
  &&{}
  + \ft12eg\,
  A_{3}{}^{\dot A\dot B}\;{\bar\chi}{}^{\dot A}\,\chi^{\dot B}
  -\ft12eg^{2}\,{\mathbb V}^{\cal \llceil MN\rrfloor,\llceil
    KL\rrfloor}\,\Theta_{\cal MN}\,\Theta_{\cal KL} 
  ~+~{\cal L}_{\mathrm{4-fermi}}  \;, 
\end{eqnarray}
where,
\begin{equation}
  \label{eq:A3}
  A_{3}{}^{\dot A \dot B}=
  \ft1{48}\,(\Gamma^{IJKL})_{\dot A\dot B}\,
  {\cal V}^{\cal MN}{}_{IJ|KL}\,\Theta_{\cal MN}  \,. 
\end{equation}
This is the Lagrangian that describes all consistent gaugings with a
constant, symmetric, embedding tensor $\Theta_\mathcal{MN}$ that
belongs to the $\mathbf{3875}+\mathbf{1}$ representation and is
subject to the quadratic constraint
$\mathcal{Q}_{\mathcal{K}\llceil\mathcal{MN}\rrfloor}=0$. 

Now consider $\Theta_{{\cal MN}}$ not as a constant tensor but as an
$x$-dependent field $\Theta_{{\cal MN}}(x)$ satisfying the
representation constraint ({\it i.e.}\ living in the ${\bf 3875}$; for
convenience we suppress the singlet representation in what follows),
but {\it not} the quadratic constraint on $\Theta_\mathcal{MN}$. To 
the Lagrangian (\ref{eq:Lgauged}) we add a new Lagrangian describing the
couplings to two-forms $B_{\mu\nu}{}^{\cal MN}$ and three-forms
$C_{\mu\nu\rho}{}^{\mathcal{K}\llceil\mathcal{MN}\rrfloor}$, 
\begin{eqnarray}
  \label{eq:LBC}
  {\cal L}_{BC} &=& -\ft18g\,\varepsilon^{\mu\nu\rho}
  B_{\mu\nu}{}^{\cal MN}\,D_{\rho}\Theta_{{\cal MN}}
  +
  \ft1{12}g^{2}\,\varepsilon^{\mu\nu\rho}\,
  C_{\mu\nu\rho}{}^{\cal K\llceil MN\rrfloor}\,
  {\cal Q}_{\mathcal{K}\llceil\mathcal{MN}\rrfloor} \;.
\end{eqnarray}
The two- and three-form potentials thus act as Lagrange multipliers to
ensure that $\Theta_{{\cal MN}}$ is constant and satisfies the
quadratic constraint.  As $\Theta_{{\cal MN}}$ is a field now, the
quadratic constraint can no longer be imposed by hand but must be
implemented in this way. 

Since the Lagrangian~(\ref{eq:Lgauged}) is supersymmetric and gauge
invariant for a constant tensor~$\Theta_{{\cal MN}}$ satisfying the
quadratic constraint, the new Lagrangian ${\cal L}_{\rm gauged}+{\cal
  L}_{BC}$ with $x$-dependent $\Theta_{{\cal MN}}$ can be made
supersymmetric and gauge invariant by introducing the proper local
transformation laws for the potentials $B_{\mu\nu}{}^{\cal MN}$ and
$C_{\mu\nu\rho}{}^{\cal K\llceil MN\rrfloor}$, while keeping
$\delta\Theta_{{\cal MN}}=0\,$.  This construction thus shows that the
supersymmetry algebra can be extended to two- and three-forms
transforming in ${\bf 3875}$ and the ${\bf 3875}+{\bf 147250}$,
respectively.  The same construction can be applied in higher
dimensions and gives a natural explanation of why in general the
$(D\!-\!1)$-forms and the $D$-forms transform in the conjugate
representations of the embedding tensor and the quadratic constraint,
respectively.

As a first exercise, we can compute the new field equation obtained
by varying the full Lagrangian with respect to\ $\Theta_{\cal MN}$.
Neglecting fermions, we find, 
\begin{eqnarray}
  \delta{\cal L}_{\rm gauged}&=& 
  -e g\left(\ft12{\cal V}^{\cal M}_{\;\;\;\;A}\,
    {\cal P}^{\mu\,A}A_{\mu}{}^{\cal N}
    +g\,{\mathbb V}^{\cal \llceil MN\rrfloor,\llceil KL\rrfloor}
    \,\Theta_{\cal KL}\right)\,\delta\Theta_{{\cal MN}}
  \nonumber\\[.5ex]
  &&{}
  -\ft14g\,\varepsilon^{\mu\nu\rho}\,A_\mu{}^{\cal M}\,
  \left(\partial_\nu A_\rho{}^{\cal N}
    +\ft23\,gX_{\cal [RS]}{}^{\cal N}\,A_\nu{}^{\cal R}
    A_\rho{}^{\cal S} \right) \,\delta\Theta_{{\cal MN}}  \;,
\end{eqnarray}
and (modulo a total derivative),
\begin{eqnarray}
  \delta{\cal L}_{BC} &=&
  \ft1{24}g\,\epsilon^{\mu\nu\rho}
  \Big(3D_{\rho}B_{\mu\nu}{}^{\cal MN}-6gZ^{\cal M}{}_{\cal
    PQ}A_{\rho}{}^{\cal N}B_{\mu\nu}{}^{\cal PQ} \nonumber\\
  &&{} \qquad \qquad +
  gY^{\cal MN}{}_{\cal K\llceil PQ\rrfloor}\,C_{\mu\nu\rho}{}^{\cal
    K\llceil PQ\rrfloor}\Big)\;\delta\Theta_{\cal MN} \;,
\end{eqnarray}
where we used the identity 
$\delta\mathcal{Q}_{\mathcal{K}\llceil\mathcal{MN}\rrfloor} =
\ft12 \delta\Theta_\mathcal{PQ}
\,Y^\mathcal{PQ}{}_{\mathcal{K}\llceil\mathcal{MN}\rrfloor}$.
Therefore the variation of the full Lagrangian ${\cal L}={\cal L}_{\rm
  gauged}+{\cal L}_{BC}$ takes the form, 
\begin{eqnarray}
  \delta{\cal L}&=&
  \ft1{24}g\,\varepsilon^{\mu\nu\rho}\,
  {\cal Y}_{\mu\nu\rho}{}^{\cal MN}\,\delta\Theta_{{\cal MN}}
  \;,
\end{eqnarray}
so that we obtain precisely the duality relation ${\cal
  Y}_{\mu\nu\rho}{}^{\cal MN}$ defined in (\ref{eq:duals}). In
particular, this shows why the scalar matrix that relates the field
strength of the $(D\!-\!1)$-forms to the embedding tensor according to
(\ref{eq:duals1}) is precisely the (non-positive definite)  matrix
$\mathbb{V}^{\llceil\mathcal{MN}\rrfloor,\llceil KL\rrfloor}$ of the
scalar potential. Clearly the analogue of this relation will hold in
any dimension.

Under general variations of vector and tensor fields, the full Lagrangian
varies as (again neglecting fermions), 
\begin{eqnarray}
  \label{varforms}
  \delta{\cal L}&=&
  -\ft14g\, \varepsilon^{\mu\nu\rho}\,\Theta_{\cal MN}\,
  \delta A_{\mu}{}^{\cal M}\,
  {\cal X}_{\nu\rho}{}^{\cal N}
  -\ft18g \,\varepsilon^{\mu\nu\rho}\,
  \left(\delta B_{\mu\nu}{}^{\cal MN}+ 2A_{[\mu}{}^{\cal M}\,\delta
    A_{\nu]}{}^{\cal N}\right) 
  \,
  D_{\rho}\Theta_{\cal MN}
  \nonumber\\[1ex]
  &&{}
  + \ft1{12}g^2\,\varepsilon^{\mu\nu\rho}\,
  \left(
    \delta C_{\mu\nu\rho}{}^{\cal K\llceil MN\rrfloor}
    + 2\,A_{\mu}{}^{\cal K}A_{\nu}{}^{\cal M}\delta A_{\rho}{}^{\cal N}
  \right)
  {\cal Q}_{\mathcal{K}\llceil\mathcal{MN}\rrfloor} \;.
\end{eqnarray} 
Thus, varying the Lagrangian with respect to all $p$-form tensor
fields and $\Theta_{\cal MN}$, one obtains the set of first order and
algebraic field equations
\begin{eqnarray} 
  \Theta_{\cal MN}\,{\cal
    X}_{\mu\nu}{}^{\cal N}&\!\!=\!\!&0\;,\quad {\cal
    Y}_{\mu\nu\rho}{}^{\cal MN} \,=\,0\;,\quad
  \partial_{\mu}\Theta_{\cal MN}\,=\,0\;,\quad 
  {\cal Q}_{\mathcal{K}\llceil\mathcal{MN}\rrfloor} \,=\,0\;, 
\end{eqnarray} 
and we recover the duality relations~${\cal
  X}^{\cal M}$ and ${\cal Y}^{\cal MN}$ that appeared in the
computation of the supersymmetry algebra (\ref{eq:dualvectors}) and
(\ref{eq:duals}), respectively.
 
Let us further remark that the full Lagrangian is invariant 
under the additional symmetry
\begin{eqnarray}
  \label{eq:dV1}
  &&\delta_\mathcal{X} {A_{\mu}}{}^{\cal M} = \xi_\mathcal{X}^{\nu}
  {\cal X}_{\nu\mu}{}^{\cal M}\;,\qquad
  \delta_\mathcal{X} B_{\mu\nu}{}^{\cal MN} =
  -2A_{[\mu}{}^{\cal \llceil M}\,\delta_\mathcal{X} A_{\nu]}{}^{\cal
  N\rrfloor}\;, 
  \nonumber\\[.5ex]
  &&
  \delta_\mathcal{X} C_{\mu\nu\rho}{}^{\cal K\llceil MN\rrfloor}
  = - 2\,A_{[\mu}{}^{\cal \llceil K} 
  A^{\vphantom{\cal M}}_{\nu}{}^{\cal \llceil M}\; 
  \delta_\mathcal{X} A_{\rho]}{}^{\cal N\rrfloor\rrfloor} \;,
\end{eqnarray}
with an arbitrary vector field $\xi_\mathcal{X}^{\nu}$. This follows 
directly from (\ref{varforms}):
\begin{eqnarray}
  \delta_\mathcal{X}{\cal L} &\propto&
  \varepsilon^{\mu\nu\rho}\,
  \Theta_{\cal MN}\,{\cal X}_{{\mu\nu}}{}^{{\cal M}}\,
  {\cal X}_{\rho\sigma}{}^{\cal N}\,\xi_\mathcal{X}^{\sigma}
  ~=~0\;.
\end{eqnarray}
Likewise, the Lagrangian is invariant under the additional symmetry
\begin{eqnarray}
  \label{eq:dV2}
  \delta_\mathcal{Y} B_{\mu\nu}{}^{\cal MN}=
  \xi_\mathcal{Y}^{\rho}\,{\cal Y}_{\rho\mu\nu}{}^{\cal MN}\;,
  \qquad
  \delta_\mathcal{Y} \Theta_{\cal MN}
  = \xi_\mathcal{Y}^{\rho}\,D_{\rho}\Theta_{\cal MN} 
\;,
\end{eqnarray}
with another arbitrary vector field $\xi_\mathcal{Y}^\mu$. 
The extra symmetries (\ref{eq:dV1}) and (\ref{eq:dV2}) are those which
have shown up already in the supersymmetry algebra and correspond to
the last two terms in (\ref{eq:susyalgebra1}).  The second one is a
standard equations-of-motion symmetry', whereas the first one is a
little more subtle as its corresponding field variations do not vanish
completely upon imposing the equations of motion.

Note that although there are of course no four-forms present in the
three-dimensional Lagrangian, their tensor gauge freedom shows up as a
shift transformation on the the three-forms (\ref{eq:C-gauge}). Since
these are the only fields transforming under this symmetry, the
Lagrangian must be invariant under the mere shift of three-forms
according to (\ref{eq:C-gauge}).  Fortunately, this invariance is
precisely ensured by the additional orthogonality (\ref{eq:QY}), showing
that the combination $C_{\mu\nu\rho}{}^{\cal K\llceil MN\rrfloor}\,
\mathcal{Q}_{\cal K\llceil MN\rrfloor}$ entering the Lagrangian is
invariant under these shifts.

A rather lengthy but straightforward calculation now shows that the
full Lagrangian ${\cal L}={\cal L}_{\rm gauged}+{\cal L}_{BC}$ is
invariant under supersymmetry provided the fields transform as
(\ref{eq:susyfermions}), (\ref{eq:susybosons}), (\ref{eq:susyB}), and
(\ref{susyC}). Here no supersymmetry variation is assigned to the
field $\Theta_{MN}$, which can still satisfy the supersymmetry
variations by virtue of the existence of the new symmetry
(\ref{eq:dV2}). Furthermore we precisely recover the new
transformation rules for the higher $p$-forms that we have derived in
section~\ref{sec:susy-algebra}.  A somewhat similar construction has
been carried out in~\cite{Bergshoeff:1996ui} to describe Roman's
massive deformation of ten-dimensional IIA
supergravity~\cite{Romans:1985tz} in terms of a nine-form potential
and an $x$-dependent parameter $m(x)$ rather than a constant
deformation parameter $m$.  What is new here is the non-trivial
representation structure of the deformation parameters and the need to
simultaneously implement on them the quadratic constraint, hence the
need for $D$-forms acting as the corresponding Lagrange multipliers.

We now return to the possible interpretation of our results, and
especially the ones of the present section, in the framework of
infinite-dimensional duality symmetries. Let us recall that the
representations found in the level decompositions of $\mathrm{E}_{11}$
\cite{Riccioni:2007au,Bergshoeff:2007qi,Bergshoeff:2007vb} are in
one-to-one correspondence with the various $p$-form fields identified
in course of our analysis and displayed in
table~\ref{tab:vector-tensor-repr}. By contrast, the embedding tensor
itself does not show up in this level decomposition, but must be added
as an `extraneous' quantity, even though it is to be treated as a
`field' in the present analysis (otherwise there would be no need for
extra $p$-form fields in the Lagrangian (\ref{eq:LBC})). In order to
better understand the link with infinite-dimensional dualities, it
would therefore be desirable to re-formulate the theory entirely in
terms of only the fields appearing in the group theoretical analysis,
and thus {\em without} $\Theta$.

At least in principle, it is possible to pass from the total
Lagrangian ${\cal L} \equiv {\cal L}_{\rm gauged} + {\cal L}_{BC}$ to
another Lagrangian which does not depend on $\Theta$, by noting that
${\cal L}$ depends on $\Theta$ at most quadratically. Accordingly, we
now regard the field equation
$\mathcal{Y}_{\mu\nu\rho}{}^\mathcal{MN}=0$ as an algebraic equation
for the (auxiliary) field $\Theta_{{\cal MN}}$, 
\begin{eqnarray}
  \label{eq:duals2}
  4\,g\,e\,\varepsilon_{\mu\nu\rho}\,{\mathbb V}^{\cal\llceil
    MN\rrfloor,\llceil KL\rrfloor}\,\Theta_{\cal KL} 
  &=&
  3\, D_{[\mu} B_{\nu\rho]}{}^\mathcal{MN}  +
  6\,A_{[\mu}{}^{\llceil\mathcal{M}}\left(\partial_{\nu}
  A_{\rho]}{}^{\mathcal{N}\rrfloor} + \ft13 g
  X_{[\mathcal{PQ}]}{}^{\mathcal{N}\rrfloor}
  A_{\nu}{}^\mathcal{P}A_{\rho]}{}^\mathcal{Q}\right) 
  \nonumber\\[.5ex]
  &&{}
  + g\,
  Y^\mathcal{MN}{}_{\mathcal{P}\llceil\mathcal{RS}\rrfloor}
  \,C_{\mu\nu\rho}{}^{\mathcal{P}\llceil\mathcal{RS}\rrfloor} 
  -6\, 
  A_{[\mu}{}^{\cal \llceil M}\,{\cal X}_{\nu\rho]}{}^{\cal N\rrfloor} 
  \;,
\end{eqnarray} 
and use it to eliminate $\Theta_{{\cal MN}}$ from the Lagrangian.
Although this equation is linear in $\Theta_{{\cal MN}}$, its solution
is rather complicated due to the hidden $\Theta$ dependence of the
tensors $X_{\mathcal{PQ}}{}^{\mathcal{N}}$,
$Y^\mathcal{MN}{}_{\mathcal{P}\mathcal{RS}}$ and
$\mathcal{X}_{\mu\nu}{}^\mathcal{M}$ on the right-hand side. 
Consequently, the solution cannot be written in closed form, but only
given as an {\em infinite series} in the $p$-forms and their
derivatives.\footnote{%%%%%%%%%%%%%%%%%%%%%%%%%%%%%%%%%%%%
  Observe that the matrix ${\mathbb V}^{\cal\llceil MN\rrfloor,\llceil 
   KL\rrfloor}$ will have zero eigenvalues at certain points of the
  scalar field configuration space.} %%%%%%%%%%%%%%%%%%%%%%%%%
We therefore exhibit only the lowest-order term of the solution
which reads
\begin{eqnarray} 
  \label{solTheta}
  \Theta_{\cal MN}&=&
  \ft34\,e^{-1}\varepsilon^{\mu\nu\rho}\, {\mathbb V}_{\cal\llceil
    MN\rrfloor,\llceil KL\rrfloor}\; \partial_{\mu}
  B_{\nu\rho}{}^\mathcal{KL} +\cdots \;.
\end{eqnarray}
Plugging (\ref{solTheta}) back into (\ref{eq:Lgauged}) and (\ref{eq:LBC}) 
we derive the bosonic kinetic term for the two-form fields in lowest
order, with the result 
\begin{eqnarray} 
  \label{Lkin3}
  {\cal L}_{{\rm kin}} &=& e\,  \partial_{[\mu}
  B_{\nu\rho]}{}^\mathcal{MN}\, \partial^{[\mu}
  B^{\nu\rho]}{}^\mathcal{KL}\; {\mathbb V}_{\cal\llceil MN\rrfloor,\llceil
  KL\rrfloor} +\dots \;, 
\end{eqnarray} 
We thus see that the inverse scalar potential matrix
${\mathbb V}_{\cal\llceil MN\rrfloor,\llceil KL\rrfloor}$ shows up as the
kinetic matrix of the $(D\!-\!1)$-forms, as would have been expected
from~(\ref{field2}). As we already pointed out above (after (\ref{eq:Pot}))
this matrix is not positive definite, unlike the kinetic matrices of 
the lower $p$-forms. Fortunately, we need to require positive definite 
kinetic terms only for those fields which carry propagating degrees of 
freedom, whence the non-positivity of the kinetic term for the 2-form 
fields in the above formula is entirely harmless.

In conclusion it is possible to re-formulate the theory in terms of
a Lagrangian that contains only the scalars and $p$-forms, but no 
embedding tensor. The price we have to pay is that the resulting
structure is rather complicated, with non-polynomial interactions 
and gauge transformations. Nevertheless, the Lagrangian obtained 
by elimination of $\Theta$ is `universal' in the sense that it 
would incorporate {\em all} gaugings, in such a way that any 
specific gauging would correspond to the 3-form
field strength $\partial_{[\mu} B_{\nu\rho]}{}^\mathcal{MN}$
acquiring a vacuum expectation value according to (\ref{solTheta}).
One may view this as a kind of `spontaneous symmetry breaking', but
of a novel kind: rather than simply breaking the rigid $\mathrm{G}$ invariance
of the original theory to some subgroup, this mechanism generates 
non-abelian gaugings from a theory with purely abelian $p$-forms 
and interactions!

By construction, the constraints on the embedding tensor exhibited and
studied in the foregoing sections must also be consistently encoded 
into this new Lagrangian. Unfortunately, due to the the 
non-polynomiality of the latter, it appears difficult to extract this
information directly and without explicit use of $\Theta$. For this
reason, it would be desirable to go beyond the mere kinematics of level
decompositions, and to `test' this non-polynomial Lagrangian (or at least 
some of its pieces, and in particular the dependence of (\ref{Lkin3}) on 
the scalars via the kinetic matrix) directly either against the 
$\mathrm{E}_{11}$ proposal of \cite{West:2001as}, or alternatively, against 
the $\mathrm{E}_{10}$ proposal of \cite{DHN,DN}. Because the latter 
admits a Lagrangian formulation (but without $D$-forms as these 
do not appear in the decomposition of $\mathrm{E}_{10}$), such tests
are possible in principle. Although this will require much more work, 
we are confident that the present results can serve as useful probes 
of M theory, or, more succinctly, of the specific proposals made 
in \cite{West:2001as} and \cite{DHN,DN}, respectively, and thereby shed new 
light on the unresolved issues with them.

%%%%%%%%%%%%%%%%%%%%%%%%%%%%%%%%%%%%%%%%%%%%%%%%%%%%%%%%%%%%%
\vspace{8mm}

\noindent
{\bf Acknowledgement}\\
\noindent
We are grateful to Eric Bergshoeff and Peter West for discussions and
correspondence. The work of H.S. is supported by the Agence Nationale
de la Recherche (ANR). The work is partly supported by EU contracts
MRTN-CT-2004-005104 and MRTN-CT-2004-512194, by INTAS contract
03-51-6346 and by NWO grant 047017015. \bigskip

%%%%%%%%%%%%%%%%%%%%%%%%%%%%%%%%%%%%%%%%%%%%%%%%%%%%%%%%%%%%
%%%%%%%%%%%%%%%%%%%%%%%%%%%%%%%%%%%%%%%%%%%%%%%%%%%%%%%%%%%%
% ---- Bibliography ----
%

%%%%%%%%%%%%%%%%%%%%%%%%%%%%%%%%%%%%%%%%%%%%%%%%%%%%%%%%%%%%%%%%%%%
%%%%%%%%%%%%%%%%%%%%%%%%%%%%%%%%%%%%%%%%%%%%%%%%%%%%%%%%%%%%%%%%%%%
%
%%%%%%%%%%%%%%%%%%%%%%%%%%%%%%%%%%%%%%%%%%%%%%%%%%%%%

\begin{thebibliography}{99}
%
\bibitem{deWit:1981}
%{deWit:1981eq}
  B. de Wit and H. Nicolai, {\it $N=8$ supergravity with local $SO(8)
    \times SU(8)$ invariance}, Phys.\ Lett.\ {\bf 108B} (1982) 285;
  %%CITATION = PHLTA,B108,285;%%
%
%\bibitem{deWit:1982ig} B. de Wit and H. Nicolai,
   {\it N=8 supergravity}, Nucl.\ Phys.\ {\bf B208} (1982) 323.
   %%CITATION = NUPHA,B208,323;%%
%
\bibitem{Gunaydin:1984qu} M.~G\"unaydin, L.~Romans and N.~Warner, {\it Gauged
 $N=8$ supergravity in five dimensions}, Phys. Lett. {\bf 154B} (1985)
 268.
%%CITATION = PHLTA,B154,268;%%
%
\bibitem{Pernici:1984xx} M.~Pernici, K.~Pilch and P.~van
  Nieuwenhuizen, {\it Gauged extended maximal supergravity in seven
    dimensions}, Phys. Lett. {\bf 143B} (1984) 103.
  %%CITATION = PHLTA,B143,103;%%
%
\bibitem{Hull}
C.M. Hull,
{\it New gaugings of N=8 supergravity}, Phys. Rev. {\bf D30} (1984)
760;
{\it Non-compact gaugings of N=8 supergravities}, Phys. Lett. {\bf
B142} (1984) 39;
{\it More gaugings of N=8 supergravity}, Phys. Lett. {\bf B148} (1984)
297. 
%{\it The minimal couplings and scalar potentials of the gauged $N=8$
%supergravities}, Class. Quantum Grav. {\bf 2} (1985) 343.
%%CITATION = PHRVA,D30,760;%%
%%CITATION = PHLTA,B142,39;%%
%%CITATION = PHLTA,B148,297;%%

\bibitem{deWit:1983gs} B.~de~Wit and H.~Nicolai, {\it The
    parallelizing $S^7$ torsion in gauged $N=8$ supergravity}, Nucl.\ 
  Phys. {\bf B231} (1984) 506.
  %%CITATION = NUPHA,B231,506;%%
%
\bibitem{cremmer}
E.~Cremmer and B.~Julia,
{\it The $N=8$ supergravity theory. I. The Lagrangian},
Phys.\ Lett.\ {\bf 80B} (1978) 48;
%%CITATION = PHLTA,B80,48;%%
{\it The SO(8) Supergravity},
  Nucl.\ Phys.\ {\bf B159} (1979) 141.
  %%CITATION = NUPHA,B159,141;%%
%
\bibitem{Nicolai:2000sc} H. Nicolai and H. Samtleben, {\it Maximal
    gauged supergravity in three dimensions}, Phys.\ Rev.\ Lett.\ {\bf
    86} (2001) 1686, {\tt hep-th/0010076};
  %%CITATION = PRLTA,86,1686;%%
%
\bibitem{Nicolai:2001sv} H.~Nicolai and H.~Samtleben, 
  {\it Compact and noncompact gauged
    maximal supergravities in three-dimensions}, JHEP {\bf 0104} (2001)
  022, {\tt hep-th/0103032}.
  %%CITATION = JHEPA,0104,022;%%
%
\bibitem{deWit:2004yr}
     B. de Wit, H. Nicolai and H. Samtleben, {\it Gauged supergravities
     in three dimensions: A panoramic overview}, Proc. 27th Johns
     Hopkins Workshop on Current Problems in Particle Theory:
     Symmetries and Mysteries of M-Theory, Goteborg, Sweden, 24-26 Aug
     2003, {\tt hep-th/0403014}. 
    %%CITATION = HEP-TH/0403014;%% 
%
\bibitem{CJS} E.~Cremmer, B.~Julia and J.~Scherk, {\it Supergravity theory 
  in 11 dimensions}, Phys.Lett. {76B} (1978) 409
%
\bibitem{deWit:2002vt} B.~de~Wit, H.~Samtleben and M.~Trigiante, {\it
    On Lagrangians and gaugings of maximal supergravities}, Nucl.\ 
  Phys.\ {\bf B655} (2003) 93, {\tt hep-th/0212239}.
  %%CITATION = NUPHA,B655,93;%%
%
\bibitem{deWit:2003hr} B.~de~Wit, H.~Samtleben and M.~Trigiante, {\it
    Gauging maximal supergravities}, Fortsch.\ Phys. {\bf 52} (2004)
  489, {\tt hep-th/0311225}.
  %%CITATION = FPYKA,52,489;%%
%
\bibitem{AndDauFerrLle} L. Andrianopoli, R. D'Auria, S. Ferrara and
  M.A. Lled\'o, {\it Gauging of flat groups in four dimensional
    supergravity}, JHEP {\bf 0207} (2002) 010, {\tt hep-th/0203206};
  {\it Duality and spontaneously broken supergravity in flat
    backgrounds}, Nucl.\ Phys.\ {\bf B640} (2002) 63, {\tt
    hep-th/0204145}.
%%CITATION = JHEPA,0207,010;%%
%%CITATION = NUPHA,B640,63;%%
%
\bibitem{Scherk:1979zr} J. Scherk and J.H. Schwarz, {\it How to Get
     Masses from Extra Dimensions}, Nucl. Phys. {\bf B153} (1979)
     61-88. 
     %%CITATION = NUPHA,B153,61;%%
%
\bibitem{deWit:2004nw} B.~de~Wit, H.~Samtleben and M.~Trigiante,
     {\it The maximal $D=5$ supergravities}, Nucl.\ Phys.\ {\bf B716}
     (2005) 215, {\tt hep-th/0412173}.
     %%CITATION = NUPHA,B716,215;%%
%
\bibitem{Samtleben:2005bp} H.~Samtleben and M.~Weidner, {\it The
       maximal $D=7$ supergravities}, Nucl.\ Phys.\ {\bf B725} (2005)
     383, {\tt hep-th/0506237}.
     %%CITATION = NUPHA,B725,383;%%
%
\bibitem{deWit:2007mt} B. de Wit, H. Samtleben M. Trigiante, {\it The
     maximal D = 4 supergravities}, JHEP, {\bf 06} (2007) 049,
     {\tt arXiv:0705.2101 [hep-th]} 
     %%CITATION = ARXIV:0705.2101;%%
%
\bibitem{Samt} E.~Bergshoeff, H.~Samtleben and E. Sezgin, {\it The 
  gaugings of maximal D=6 supergravity}, MIFP-07-33, 
       {\tt arXiv:0712.4277 [hep-th]} 
     %%CITATION = ARXIV:0712.4277;%%
%
\bibitem{Hull2} C.M.\ Hull, {\it New gauged $N=8$, $D=4$
       supergravities}, Class. Quantum Gravity {\bf 20} (2003) 5407,
     {\tt hep-th/0204156}.
%%CITATION = HEP-TH 0204156;%%
%
\bibitem{deWit:2003hq} B.~de~Wit, H.~Samtleben and M.~Trigiante, {\it Maximal
    supergravity from IIB flux compactifications}, Phys.\ Lett. {\bf 583B}
  (2004) 338, {\tt hep-th/0311224}.
%%CITATION = PHLTA,B583,338;%%
%
\bibitem{Dall'Agata:2005ff}
  G.~Dall'Agata and S.~Ferrara,
  {\it Gauged supergravity algebras from twisted tori compactifications with
  fluxes},
  Nucl.\ Phys.\  {\bf B717} (2005) 223,
  {\tt hep-th/0502066}.
  %%CITATION = NUPHA,B717,223;%%
%
\bibitem{Andrianopoli:2005jv}
 L.~Andrianopoli, M.A.~Lledo and M.~Trigiante,
 {\it The Scherk-Schwarz mechanism as a flux compactification with internal
 torsion},
  JHEP {\bf 05} (2005) 051, {\tt hep-th/0502083}. 
  %%CITATION = JHEPA,0505,051;%%
  %%CITATION = HEP-TH/0701218;%%
%
\bibitem{D'Auria:2005}
  R.~D'Auria, S.~Ferrara and M.~Trigiante,
 {\it E$_{7(7)}$ symmetry and dual gauge algebra of M-theory on a twisted
 seven-torus},
  Nucl.\ Phys.\  {\bf B732} (2006) 389,
  {\tt hep-th/0504108};
  %%CITATION = NUPHA,B732,389;%%
%
%  R.~D'Auria, S.~Ferrara and M.~Trigiante,
  {\it Supersymmetric completion of M-theory 4D-gauge algebra from
  twisted  tori and fluxes},
  JHEP {\bf 0601} (2006) 081, {\tt hep-th/0511158}.
  %%CITATION = JHEPA,0601,081;%%
%
\bibitem{Schon:2006kz} J.~Sch\"on and M.~Weidner, {\it Gauged $N = 4$
    supergravities}, JHEP {\bf 05} (2006) 034, {\tt hep-th/0602024}.
  %%CITATION = JHEPA,0605,034;%%
%
\bibitem{Hull:2006tp}
  C.~M.~Hull and R.~A.~Reid-Edwards,
{\it Flux compactifications of M-theory on twisted tori},
  JHEP {\bf 10} (2006) 086,
  {\tt hep-th/0603094}.
  %%CITATION = JHEPA,0610,086;%%
%
\bibitem{deVroome:2007zd} M.~de Vroome and B.~de Wit, {\it Lagrangians
    with electric and magnetic charges in $N=2$ supersymmetric gauge
    theories}, JHEP {\bf 08} (2007) 064, {\tt arXiv:0707.2717
    [hep-th]}.
    %%CITATION = ARXIV:0707.2717;%%
%
\bibitem{Dall'Agata:2007sr} G. Dall'Agata, N. Prezas, H. Samtleben and
                  M. Trigiante, {\it Gauged Supergravities from
                  Twisted Doubled Tori and Non-Geometric String
                  Backgrounds}, {\tt arXiv:0712.1026 [hep-th]}. 
     %%CITATION = ARXIV:0712.1026;%%
%
\bibitem{Derendinger:2007xp} J.P.~Derendinger,
                P.M.~Petropoulos and N.~Prezas, {\it Axionic symmetry
                  gaugings in N = 4 supergravities and their
                  higher-dimensional origin}, Nucl.\ Phys.\ {\bf B785}
                (2007) 115, {\tt arXiv:0705.0008 [hep-th]}.
 %%CITATION = NUPHA,B785,115;%%

\bibitem{SamtlebenWeidner2} H.~Samtleben and M.~Weidner, {\it Gauging
    hidden symmetries in two dimensions}, JHEP {\bf 08} (2007) 076,
    {\tt arXiv:0705.2606 [hep-th]}. 
    %%CITATION = ARXIV:0705.2606;%%
%
  \bibitem{deWit:2005hv} B.~de~Wit and H.~Samtleben, {\it Gauged
      maximal supergravities and hierarchies of nonabelian
      vector-tensor systems}, { Fortsch. Phys.} {\bf 53} (2005) 442,
    {\tt hep-th/0501243}.
%%CITATION = FPYKA,53,442;%%
%
 \bibitem{West:2004kb} P.C. West, {\it E(11) origin of brane
  charges and U-duality multiplets}, JHEP {\bf 08} (2004) 052, {\tt
  hep-th/0406150}.
     %%CITATION = HEP-TH/0406150;%%
%
\bibitem{Riccioni:2007au} F. Riccioni and P. West, {\it The E(11)
     origin of all maximal supergravities} JHEP {\bf 07} (2007) 063,
     {\tt arXiv:0705.0752 [hep-th]}. 
     %%CITATION = ARXIV:0705.0752;%%
%
\bibitem{Bergshoeff:2007qi} E.A. Bergshoeff, I. De Baetselier and
                  T.A. Nutma, {\it E(11) and the embedding tensor},
                  JHEP {\bf 09} (2007) 047, {\tt arXiv:0705.1304
                  [hep-th]}. 
     %%CITATION = ARXIV:0705.1304;%%
%
\bibitem{Bergshoeff:2007vb} E.A. Bergshoeff, J. Gomis, T.A. Nutma and
     D. Roest, {\it Kac-Moody Spectrum of (Half-)Maximal
     Supergravities}, {\tt arXiv:0711.2035 [hep-th]}.
     %%CITATION = ARXIV:0711.2035;%%
%
\bibitem{West:2001as}  P.C. West, {\it E(11) and M theory},
   Class. Quant. Grav. {\bf 18} (2001) 4443, {\tt hep-th/0104081}.
     %%CITATION = HEP-TH/0104081;%%
%
\bibitem{Schnakenburg:2001ya} I. Schnakenburg and P.C. West,
   {\it Kac-Moody symmetries of IIB supergravity}, Phys. Lett. {\bf
   B517} (2001) 421, {\tt hep-th/0107181}.
     %%CITATION = HEP-TH/0107181;%%
%
\bibitem{Elitzur:1997zn} S. Elitzur, A. Giveon, D. Kutasov and
                  E. Rabinovici, {\it Algebraic aspects of matrix
                  theory on $T^d$}, Nucl. Phys. {\bf B509} (1998)
                  122-144, {\tt hep-th/9707217}.
     %%CITATION = HEP-TH/9707217;%%
%
\bibitem{Obers:1998fb} N.A. Obers and B. Pioline, {\it U-duality and
     M-theory}, Phys. Rept. {\bf 318} (1999) 113-225", {\tt
     hep-th/9809039}. 
     %%CITATION = HEP-TH/9809039;%% 
%
\bibitem{Iqbal:2001ye} A. Iqbal, A. Neitzke and C. Vafa, {\it A
     mysterious duality}, Adv. Theor. Math. Phys. {\bf 5} (2002)
     769-808, {\tt hep-th/0111068}. 
     %%CITATION = HEP-TH/0111068;%%
%
\bibitem{Romans:1985tz} L.J. Romans, {\it Massive N=2a Supergravity
       in Ten-Dimensions}, Phys. Lett. {\bf B169} (1986) 374.
  %%CITATION = PHLTA,B169,374;%%
%
   \bibitem{deWit:2005ub} B.~de~Wit, H.~Samtleben and M.~Trigiante, {\it
       Magnetic charges in local field theory}, JHEP {\bf 09} (2005)
     016, {\tt hep-th/0507289}.
     %%CITATION = JHEPA,0509,016;%%
%
\bibitem{LeCoLi92} M.~van Leeuwen, A.~Cohen, and B.~Lisser, {{LiE}, a
    computer algebra package for {L}ie group computations}, {Computer
    Algebra Nederland, Amsterdam} (1992).
%
\bibitem{deWit:2000wu} B. de Wit and H. Nicolai, {\it Hidden
     symmetries, central charges and all that},
     Class. Quant. Grav. {\bf 18} (2001) 3095-3112, {\tt hep-th/0011239}.
     %%CITATION = HEP-TH/0011239;%% 
%    
   \bibitem{deWit:1988ig} B.~de Wit, J.~Hoppe and H.~Nicolai, {\it On
       the quantum mechanics of supermembranes}, Nucl. Phys. {\bf B305}
     (1988) 545. 
     %%CITATION = NUPHA,B305,545;%%
%
\bibitem{Banks:1996vh} T.~Banks, W.~Fischler, S.H.~Shenker and
  L.~Susskind, {\it M-Theory as a matrix model: A conjecture}, Phys.
  Rev. {\bf D55} (1997) 5112, {\tt hep-th/9610043}.
  %%CITATION = HEP-TH/9610043;%%
% 
\bibitem{Hull:1994ys} C.~Hull and P.K.~Townsend, {\it Unity of
    superstring dualities}, Nucl. Phys. {\bf B438} (1995) 109, {\tt
    hep-th/9410167}.
  %%CITATION = HEP-TH/9410167;%%
%  
\bibitem{Englert:2007qb} F.~Englert, L.~Houart, A.~Kleinschmidt,
  H.~Nicolai and T.~Nassiba, {\it An $E_9$ multiplet of BPS states},
  JHEP {\bf 05} (2007) 065, {\tt hep-th/0703285}.
  %%CITATION = HEP-TH/0703285;%%
%
\bibitem{Riccioni:2007ni} F. Riccioni and P. West, {\it E(11)-extended
     spacetime and gauged supergravities}, {\tt arXiv:0712.1795
     [hep-th]}. 
     %%CITATION = ARXIV:0712.1795;%%
%
\bibitem{KNS} K.~Koepsell, H.~Nicolai and H.~Samtleben, {\it On the 
  Yangian quantum symmetry of maximal supergravity in two dimensions},
  JHEP {\bf 04} (1999) 023, {\tt hep-th/9903111}.
%%CITATION = HEP-TH 9903111;%%.
%
\bibitem{Bergshoeff:1996ui} E.~Bergshoeff, M.~de Roo, M.B.~Green,
  G.~Papadopoulos and P.K.~Townsend, {\it Duality of Type II 7-branes
  and 8-branes},  Nucl.\ Phys.\ {\bf B470} (1996) 113,
  {\tt hep-th/9601150}.
 %%CITATION = NUPHA,B470,113;%%
%
%\bibitem{West} P.C.~West, {\it $E_{11}$ and M theory}, Class. Quant. 
%   Grav. {\bf 18} (2001) 4443, {\tt hep-th/0104081}.
%
\bibitem{DHN} T.~Damour, M.~Henneaux and H.~Nicolai, {\it $E_{10}$
  and a small tension expansion of M theory}, Phys. Rev. Lett. {\bf 89}
  (2002) 221601, {\tt hep-th/0207267}.
%
\bibitem{DN} T.~Damour and H.~Nicolai, {\it Eleven dimensional
    supergravity and the $E_{10}/K(E_{10})$ $\sigma$-model at low
    $A_9$ levels}, in: Group Theoretical Methods in Physics, Institute
    of Physics Conference Series No. 185, IoPP (2005), {\tt
    hep-th/0410245}. 
%%%%%%%%%%%%%%%%%%%%%%%
  %
%%%%%%%%%%%%%%%%%%%%%%%%%%%%%%%%%%%%%%%%%%%%%%%%%%%%%%%%%%%%%%%%%%%
\end{thebibliography}
\end{document}